%% file: Literature.tex
\documentclass[entropy,review,submit,manyauthors,dvipdfm,12pt,a4paper]{mdpi}

\lastpage{x}
\doinum{10.3390/------}
\pubvolume{xx}
\pubyear{2010}
\history{Received: xx / Accepted: xx / Published: xx}

\usepackage{amssymb,amsmath}
\usepackage{graphicx}
\usepackage{subfigure}
\usepackage{wrapfig}
\usepackage{caption}
\Title{Using Quantum Computers for Quantum Simulation}
\Author{Katherine L. Brown $^{1}$\thanks{Corresponding author: pyklb@leeds.ac.uk}, William J. Munro $^{2,3}$ and Vivien M. Kendon $^{1}$}
\address{$^{1}$ School of Physics and Astronomy, University of Leeds, Woodhouse Lane, Leeds LS2 9JT, UK \\
$^{2}$ National Institute of Informatics, 2-1-2 Hitotsubashi, Chiyoda-ku, Tokyo, 101-8430, Japan\\
$^{3}$ NTT Basic Research Laboratories, NTT Corporation, 3-1 Morinosato-Wakamiya, Atsugi-shi, Kanagawa-ken 243-0198, Japan}

\abstract{Numerical simulation of quantum systems is crucial to further our
understanding of natural phenomena.  Many systems of key interest and 
importance, in areas such as superconducting materials and quantum chemistry, 
are thought to be described by models which we cannot solve with
sufficient accuracy, neither analytically nor numerically
with classical computers.
Using a quantum computer to simulate such quantum systems has been viewed as 
a key application of quantum computation from the very beginning of
the field in the 1980s.  Moreover, useful results beyond the reach of
classical computation are expected to be accessible with fewer than a
hundred qubits, making quantum simulation potentially one of the
earliest practical applications of quantum computers.  In this paper
we survey the theoretical and experimental development
of quantum simulation using quantum computers, from the first ideas
to the intense research efforts currently underway.}
\begin{document}
\tableofcontents

\part{The Theory Behind Quantum Simulation}\label{Part1}
\input{Intro}
\input{Universal}
\input{Measure}

\input{Initialise}

\input{Evolve}

\input{FB}

\part{Experimental Implementation of Quantum Simulations}\label{Part2}
\input{Over}

\input{Proof}

\input{OL}

\input{Electrons}

\input{Outlook}

\section*{Acknowledgments}
We thank Clare Horsman for careful reading of the manuscript.
KLB is supported by a UK EPSRC CASE studentship from Hewlett Packard.
VMK is funded by a UK Royal Society University Research Fellowship.
WJM acknowledges part support from MEXT in Japan.

\bibliographystyle{plainnat}
\bibliography{bibliography,bibextra}
\end{document}

%% file: Intro.tex
\section{Introduction}\label{Intro}

The role of numerical simulation in science is to work out in detail what
our mathematical models of physical systems predict.  When the models
become too difficult to solve by analytical techniques, or details are
required for specific values of parameters, numerical computation
can often fill the gap.  This is only a practical option if the calculations
required can be done efficiently with the resources available.
As most computational scientists know well, many calculations we
would like to do require more computational power than we have.  
Running out of computational power is nearly ubiquitous
whatever you are working on, but for those working on quantum
systems this happens for rather small system sizes.  Consequently,
there are significant open problems in important areas, such as
high temperature superconductivity, where progress is slow because we
cannot adequately test our models or use them to make predictions.

Simulating a fully general quantum system on a 
classical computer is possible only for very small systems,
because of the exponential scaling of the Hilbert space
with the size of the quantum system.
To appreciate just how quickly this takes us beyond reasonable
computational resources, consider the classical memory
required to store a fully general state $|\psi_n\rangle$ of
$n$ qubits (two-state quantum systems).  The Hilbert space for
$n$ qubits is spanned by $2^n$ orthogonal states, labeled $|j\rangle$
with $0\le j < 2^n$.  The $n$ qubits
can be in a superposition of all of them in different proportions,
\begin{equation}
|\psi_n\rangle = \sum_{j=0}^{2^n-1} c_j|j\rangle
\label{eq:qstate}
\end{equation}
To store this description of the state in a classical computer,
we need to store all of the complex numbers $\{c_j\}$.  Each requires
two floating point numbers (real and imaginary parts).  
Using 32 bits (4 bytes) for each floating point number, a quantum state
of $n=27$ qubits will require 1 Gbyte of memory -- a new desktop
computer in 2010 probably has around 2 to 4 Gbyte of memory in total. 
Each additional qubit doubles the memory, so 37 qubits would
need a Terabyte of memory -- a new desktop computer in 2010 probably has a
hard disk of this size.  The time that would be required to
perform any useful calculation on this size of data is actually what
becomes the limiting factor.  One of the largest simulations of qubits
on record \cite{deraedt06a} computed the evolution of 36 qubits in a
quantum register
using one Terabyte of memory, with multiple computers for the processing.
Simulating more than 40 qubits in a fully general superposition state
is thus well beyond our current capabilities.  Computational physicists can
handle larger systems if the model restricts the dynamics to only part of
the full Hilbert space.
Appropriately designed methods then allow larger classical simulations
to be performed \citep{verstraete04a}.  
However, any model is only as good as its assumptions, and capping the
size of the accessible part of the
Hilbert space below $2^{36}$ orthogonal states for all system
sizes is a severe restriction.

The genius of Feynman in 1982 was to come up with an idea for how
to circumvent the difficulties of simulating quantum systems classically 
\cite{Feynman1982}.
The enormous Hilbert space of a general quantum state
can be encoded and stored efficiently on a quantum computer using
the superpositions it has naturally.
This was the original inspiration for quantum computation,
independently proposed also by Deutsch \cite{deutsch85a} a few years
later.
The low threshold for useful quantum simulations, of upwards of
36 or so qubits, means it is widely expected to be
one of the first practical applications of a quantum 
computer.
Compared to the millions of qubits needed for useful instances of other
quantum algorithms, such as Shor's algorithm for factoring \cite{Shor1997},
this is a realistic goal for current experimental research to work towards.
We will consider the experimental challenges in the latter sections of
this review, after we have laid out the theoretical requirements.

Although a quantum computer can efficiently store the quantum state
under study, it is not a ``drop in'' replacement for a classical
computer as far as the methods and results are concerned.
A classical simulation of a quantum system gives us access to the
full quantum state, i.e., all the $2^n$ complex numbers $\{c_j\}$ in equation
(\ref{eq:qstate}).  A quantum computer storing the same quantum state
can in principle tell us no more than whether one of the $\{c_j\}$ is
non-zero, if we directly measure the quantum state in the computational
basis.  As with all types of quantum algorithm, an extra step is required
in the processing to concentrate the information we want into the
register for the final measurement.  Particularly for quantum simulation, 
amassing enough useful information also typically requires a significant
number of repetitions of the simulation.  
Classical simulations of quantum systems are usually ``strong simulations''
\cite{vandennest08a,vandennest09a}
which provide the whole probability distribution,
and we often need at least a significant part of this, e.g., for
correlation functions, from a quantum simulation.
If we ask only for sampling from the probability distribution,
a ``weak simulation'',
then a wider class of quantum computations can be simulated
efficiently classically, but may require repetition to provide
useful results, just as the quantum computation would.
Clearly, it is only worth using a quantum computer when neither
strong nor weak simulation can be performed efficiently classically, and
these are the cases we are interested in for this review.

As with all quantum algorithms, the three main steps, initialization,
quantum processing, and data extraction (measurement) must all be
performed efficiently to obtain a computation that is efficient overall.
Efficient in this context will be taken to mean
using resources that scale polynomially
in the size of the problem, although this isn't always a reliable
guide to what can be achieved in practice.
For many quantum algorithms, the initial state of the computer is
a simple and easy to prepare state, such as all qubits set to zero.
However, for a typical quantum simulation, the initial state we want is often
an unknown state that we are trying to find or characterise, such as
the lowest energy state.  The special techniques required to deal with this
are discussed in section \ref{Initialisation}.
The second step is usually the time evolution of the 
Hamiltonian.
Classical simulations use a wide variety of methods, depending
on the model being used, and the information being calculated.
The same is true for quantum simulation, although the diversity is
less developed, since we don't have the possibility to actually
use the proposed methods on real problems yet and refine through practice.
Significant inovation in classical simulation methods arose as
a response to practical problems encountered when theoretical
methods were put to the test, and we can expect the same to
happen with quantum simulation.
The main approach to time evolution using a universal quantum
computer is described in section \ref{Lloyd}, in which
the Lloyd method for evolving the Hamiltonian using Trotterization
is described.
In section \ref{evolve}, further techniques are described, including
the quantum version of the pseudo-spectral method that 
converts between position and momentum space to evaluate different
terms in the Hamiltonian using the simplest representation for each, and
quantum lattice gases, which can be used as a general differential
equation solver in the same way that classical lattice gas and
lattice Boltzmann methods are applied.
It is also possible to take a
direct approach, in which the Hamiltonian of the quantum simulator
is controlled in such a way that it behaves like 
the one under study -- an idea already well established in the 
Nuclear Magnetic Resonance (NMR) community. 
The relevant theory is covered in section \ref{UniversalH}.
The final step is data extraction.  Of course, data extraction methods
are dictated by what we want to calculate, and this in turn affects
the design of the whole algorithm, which is why it is most naturally
discussed before initialization, in section \ref{Measure}.

For classical simulation, we rarely use anything other than standard
digital computers today.  Whatever the problem, we map it onto the registers 
and standard gate operations available in a commercial computer (with the
help of high level programming languages and compilers).
The same approach to quantum simulation makes use of the 
quantum computer architectures proposed for universal quantum computation.
The seminal work of Lloyd \cite{Lloyd1996} gives the conditions under
which quantum simulations can be performed efficiently on a universal
quantum computer.  The subsequent development of quantum simulation
algorithms for general purpose quantum computers
accounts for a major fraction of the theoretical work in quantum simulation.
However, special purpose computational modules are still used for classical 
applications in many areas, such as fast real time control of
experimental equipment, or video rendering on graphics cards to control 
displays, or even mundane tasks such as controlling a toaster, or
in a digital alarm clock.
A similar approach can also be used for quantum simulation.
A quantum simulator is a device which is designed to simulate
a particular Hamiltonian, and it may not be 
capable of universal quantum computation.  Nonetheless,
a special purpose quantum simulator could still be fast and efficient
for the particular simulation it is built for. This would allow a useful
device to be constructed before we have the technology for universal
quantum computers capable of the same computation.
This is thus a very active area of current research.
We describe a selection of these in the experimental sections \ref{proof}
to \ref{OA},
which begins with its own overview in section \ref{over}.

While we deal here strictly with quantum simulation of quantum systems,
some of the methods described here, such as lattice gas automata,
are applicable to a wider class of problems, which will be mentioned
as appropriate.
A short review such as this must necessarily be brief and selective in
the material covered from this broad and active field of research.
In particular, the development of Hamiltonian simulation applied to 
quantum algorithms begun by the seminal work of Aharonov and Ta-Shma
\cite{Aharonov2003} -- which is worthy of a review in itself -- is
discussed only where there are implications for practical applications.
Where choices had to be made, the focus has been on relevance to
practical implementation for solving real problems, and reference has been
made to more detailed reviews of specific topics, where they already exist.
The pace of development in this exciting field is such that it will
in any case be important to refer to more recent publications to
obtain a fully up to date picture of what has been achieved.

%% file: Universal.tex
\section{Universal Quantum Simulation} \label{Universal}

The core processing task in quantum simulation will usually
be the time evolution of a quantum system under a given Hamiltonian,
\begin{equation}
|\Psi(t)\rangle = \exp(i\hat{H}t)|\Psi(0)\rangle
\end{equation}
Given the initial state $|\Psi(0)\rangle$ and the Hamiltonian
$\hat{H}$, which may itself be time dependent, calculate the
state of the system $|\Psi(t)\rangle$ at time $t$.
In many cases it is the properties of a system governed by
the particular Hamiltonian that are being sought, and pure quantum
evolution is sufficient. 
For open quantum systems where coupling to another system or
environment plays a role, the appropriate master equation 
will be used instead.
In this section we will explore how accomplish the time evolution
of a Hamiltonian efficiently, thereby explaining the basic
theory underlying quantum simulation.

\subsection{Lloyd's method}\label{Lloyd}

Feynman's seminal ideas \cite{Feynman1982} from 1982
were fleshed out by Lloyd in 1996,
in his paper on universal quantum simulators \cite{Lloyd1996}.
While a quantum computer can clearly store the quantum state
efficiently compared with a classical computer, this is only
half the problem.  It is also crucial that the computation
on this superposition state can be performed efficiently, the
more so because classically we actually run out of computational
power before we run out of memory to store the state.
Lloyd notes that simply by turning on and off the correct
sequence of Hamiltonians, a system can be made to evolve
according to any unitary operator.
By decomposing the unitary operator into a sequence of standard quantum gates,
Vartiainen et al \cite{Vartiainen2004} provide a method for doing
this with a gate model quantum computer.
However, an arbitrary unitary operator requires exponentially
many parameters to specify it, so we don't get an efficient
algorithm overall.  A unitary operator with
an exponential number of parameters requires exponential
resources to simulate it in both the quantum and classical cases.
Fortunately, (as Feynman had envisaged), any system that is consistent with 
special and general relativity evolves according to local interactions.
All Hamiltonian evolutions $\hat{H}$ with only local interactions
can be written in the form
\begin{equation}
\hat{H} = \sum^n_{j=1}\hat{H}_{j}
\label{eq:Hj}
\end{equation} 
where each of the $n$ local Hamiltonians $\hat{H}_j$
acts on a limited space containing at most $\ell$ 
of the total of $N$ variables.  By ``local'' we only require
that $\ell$ remains fixed as $N$ increases,
we don't require that the $\ell$ variables are actually
spatially localised, allowing efficient simulation for
many non-relativistic models with long-range interactions.
The number of possible distinct terms
$\hat{H}_j$ in the decomposition of $\hat{H}$ is given by the
binomial coefficient ${N \choose \ell} < N^\ell/\ell!$.
Thus $n < N^\ell/\ell!$ is polynomial in $N$.
This is a generous upper bound in many practical cases: for Hamiltonians
in which each system interacts with at most $\ell$ nearest neighbours,
$n\simeq N$.

In the same way that classical simulation of the time evolution 
of dynamical systems is often performed, the total
simulation time $t$ can be divided up into $\tau$ small discrete steps.
Each step is approximated using a Trotter-Suzuki
\cite{Trotter1959,Suzuki1993} formula,
\begin{equation}
\exp\{i\hat{H}t\} = \left(\exp\{i\hat{H}_1t/\tau\} \dots \exp\{i\hat{H}_n t/\tau\}\right)^\tau
+ \sum_{j'> j}[H_{j'}, H_j]t^2/2\tau + \text{higher order terms}
\label{eq:TS}
\end{equation}
The higher order term of order $k$ is bounded by
$\tau||\hat{H}t/\tau||_\text{sup}^k/k!$, where $|| \hat{A} ||_{sup}$
is the supremum, or maximum expectation value,
of the operator $\hat{A}$ over the states of interest.
The total error is less than 
$||\tau\{\exp(i\hat{H}t/\tau) - 1 -i\hat{H}t/\tau\}||_\text{sup}$ if just the
first term in equation (\ref{eq:TS}) is used to approximate $\exp(i\hat{H}t)$.
By taking $\tau$ to be sufficiently large the error can be made as
small as required.  For a given error $\epsilon$,
from the second term in equation (\ref{eq:TS}) we have
$\epsilon \propto t^2/\tau$.  A first order Trotter-Suzuki simulation
thus requires $\tau\propto t^2/\epsilon$.

Now we can check that the simulation scales efficiently in the number
of operations required.
The size of the most general Hamiltonian $\hat{H}_j$ between $\ell$
variables depends on the
dimensions of the individual variables but will be bounded by a
maximum size $g$.  The Hamiltonians $\hat{H}$ and $\{\hat{H}_j\}$
can be time dependent so long as $g$ remains fixed.
Simulating $\exp\{i\hat{H}_jt/\tau\}$ requires $g_j^2$ operations
where $g_j\le g$ is the dimension of the variables involved in $\hat{H}_j$.
In equation (\ref{eq:TS}), each local operator $\hat{H}_j$ is
simulated $\tau$ times.  Therefore, the total number of operations
required for simulating $\exp\{i\hat{H}t\}$ is bounded by $\tau n g^2$.
Using $\tau\propto t^2/\epsilon$, the number of operations $Op_{\text{Lloyd}}$
is given by
\begin{equation}
Op_{\text{Lloyd}} \propto t^2 n g^2/\epsilon
\label{eq:lloydscale}
\end{equation}
The only dependence on the system size $N$ is in $n$,
and we already determined that $n$ is
polynomial in $N$, so the number of operations is indeed
efficient by the criterion of polynomial scaling in the problem size.

The simulation method provided by Lloyd that we just described
is straightforward but very general. Lloyd laid the groundwork for 
subsequent quantum simulation development, by providing 
conditions (local Hamiltonians) under which it will be possible in theory to
carry out efficient quantum simulation, and describing an
explicit method for doing this.
After some further consideration of the way the errors scale,
the remainder of this section elaborates on
exactly which Hamiltonians $\hat{H}_q$ in a quantum simulator can efficiently
simulate which other Hamiltonians $\hat{H}_j$ in the system under study.

\subsection{Errors and efficiency}\label{noise}

Although Lloyd \cite{Lloyd1996} explicitly notes that to keep
the total error below $\epsilon$, each operation must 
have error less than $\epsilon/(\tau n g^2)$ where $n = \text{poly}(N)$,
he does not discuss the implications of this scaling as an
inverse polynomial in $N$.
For digital computation, we can improve the accuracy of our
results by increasing the number of bits of precision
we use.  In turn, this increases the number of
elementary (bitwise) operations required to process the data.
To keep the errors below a chosen $\epsilon$ by the end of the computation,
we must have $\log_2(1/\epsilon)$ accurate bits in our output register.
The resources required to achieve this in an efficient computation
scale polynomially in $\log_2(1/\epsilon)$.
In contrast, as already noted in equation (\ref{eq:lloydscale}),
the resources required 
for quantum simulation are proportional to $t^2 n g^2/\epsilon$, so
the dependence on $\epsilon$ is inverse, rather than log inverse.

The consequences of this were first discussed by
Brown et al \cite{Brown2006}, who point out that
all the work on error correction for quantum computation
assumes a logarithmic scaling of errors with the size of the computation,
and they experimentally verify that the errors do indeed scale inversely for an
NMR implementation of quantum simulation.
To correct these errors thus requires exponentially more operations for
quantum simulation than for a typical (binary encoded) quantum
computation of similar size and precision.
This is potentially a major issue, once quantum simulations
reach large enough sizes to solve useful problems.  
The time efficiency of the computation for any quantum simulation method
will be worsened due to the error correction overheads.
This problem is mitigated somewhat because we may not actually need
such high precision for quantum simulation as we do for calculations
involving integers, for example.
However, Clark et al \cite{Clark2008}
conducted a resource analysis for a quantum simulation to
find the ground state energy of the transverse Ising model performed
on a circuit model quantum computer. They found that,
even with modest precision, error correction requirements result in
unfeasibly long simulations for systems that would be possible to
simulate if error correction weren't necessary.
One of the main reasons for this is the use of Trotterization,
which entails a large number of steps $\tau$
each composed of many operations with associated imperfections
requiring error correction.

Another consequence of the polynomial scaling of the errors,
explored by Kendon et al \cite{Kendon2010}, is that analogue
(continuous variable) quantum computers may be equally suitable for quantum 
simulation, since they have this same error scaling for any computation
they perform.  This means they are usually considered only for small
processing tasks as part of quantum communications networks, where
the poor scaling is less of a problem.  As Lloyd notes \cite{Lloyd1996},
the same methods as for discrete systems generalise directly 
onto continuous variable systems and Hamiltonians.

On the other hand, this analysis doesn't include potential savings that
can be made when implementing the Lloyd method, 
such as by using parallel processing to
compute simultaneously the terms in equation (\ref{eq:Hj}) that commute.  
The errors due to decoherence can also be exploited to simulate
the effects of noise on the system being studied, see section
\ref{open}.
Nonetheless, the unfavorable scaling of the error correction requirements
with system size in quantum simulation remains an
under-appreciated issue for all implementation methods.

\subsection{Universal Hamiltonians}\label{UniversalH}

Once Lloyd had shown that quantum simulation can be done efficiently overall,
attention turned to the explicit forms of the Hamiltonians,
both the $\{\hat{H}_j\}$ in the system to be simulated, and
the $\{\hat{H}_q\}$ available in the quantum computer.
Since universal quantum computation amounts to being able to
implement any unitary operation on the register of the quantum
computer, this includes quantum simulation as a special case,
i.e., the unitary operations derived from local Hamiltonians.
Universal quantum computation is thus sufficient for quantum
simulation, but this leaves open the possibility that
universal quantum simulation could be performed 
equally efficiently with less powerful resources.
There is also the important question of how much the efficiency can
be improved by exploiting the $\{\hat{H}_q\}$ in the
quantum computer directly rather than through standard quantum gates.

The natural idea that mapping directly between the $\{\hat{H}_j\}$
and the $\{\hat{H}_q\}$ should be the most efficient way to do 
quantum simulation resulted in a decade of research that has
answered almost all the theoretical questions one can ask about
exactly which Hamiltonians can simulate which
other Hamiltonians, and how efficiently.
The physically-motivated setting for much of this work is a
quantum computer with a single, fixed 
interaction between the qubits, that can be turned on and off but
not otherwise varied, along with arbitrary local control operations
on each individual qubit.
This is a reasonable abstraction of a typical quantum computer architecture:
controlled interactions between qubits are usually
hard and/or slow compared with rotating individual qubits.
Since most non-trivial interaction Hamiltonians can be used to
do universal quantum computation, it follows they can generally
simulate all others (of the same system size or smaller) as well.
However, determining the optimal control sequences and resulting efficiency
is computationally hard in the general case
\cite{Wocjan2001,Wocjan2001a,Wocjan2002}, which is not
so practical for building actual universal quantum simulators.
These results are thus important for the theoretical 
understanding of the interconvertability of Hamiltonians,
but for actual simulator design we will need to choose Hamiltonians
$\{\hat{H}_q\}$
for which the control sequences can be obtained efficiently.

Dodd et al \cite{Dodd2002}, Bremner et al \cite{Bremner2002},
and Nielsen et al \cite{Nielsen2002} characterised 
non-trivial Hamiltonians as entangling Hamiltonians,
in which every subsystem is coupled to every 
other subsystem either directly or via intermediate subsystems.
When the subsystems are qubits (two-state quantum systems),
multi-qubit Hamiltonians involving an even number of
qubits provide universal simulation,
when combined with local unitary operations.
Qubit Hamiltonians where the terms
couple only odd numbers of qubits are 
universal for the simulation of one fewer logical qubits (using a special
encoding) \cite{Bremner2004}.
When the subsystems are qudits (quantum systems of dimension $d$),
any two-body qudit entangling Hamiltonian is universal,
and efficiently so, when combined with local unitary operators
\cite{Nielsen2002}.
This is a useful and illuminating approach because of the fundamental
role played by entanglement in quantum information processing.  
Entanglement can only be generated by interaction (direct or indirect)
between two (or more) parties.  The local unitaries and controls
can thus only move the entanglement around, they cannot increase it.
These results show that generating enough entanglement can be completely
separated from the task of shaping the exact form of the Hamiltonian.
Further work on general Hamiltonian simulation has been done 
by McKague et al \cite{McKague2009}
who have shown how to conduct a multipartite simulation 
using just a real Hilbert space.  While not of practical
importance, this is significant in relation to foundational
questions.  It follows from their work that Bell inequalities
can be violated by quantum states restricted to a real Hilbert space.
Very recent work by Childs et al \cite{Childs2010a} fills in most of
the remaining gaps in our knowledge of the conditions under which
two-qubit Hamiltonians are universal for approximating other
Hamiltonians (equally applicable to both quantum simulation and computation).
There are only three special types of two-qubit Hamiltonians
that aren't universal for simulating other two-qubit Hamiltonians, and
some of these are still universal for simulating Hamiltonians on more than
two qubits.

\subsection{Efficient Hamiltonian simulation}\label{EfficientH}

The other important question about using one Hamiltonian
to simulate another is how efficiently it can be done.
The Lloyd method described in section \ref{Lloyd} can be
improved to bring the scaling with $t$ down from quadratic,
equation (\ref{eq:lloydscale}), to close to linear by using
higher order terms from the Trotter-Suzuki expansion \cite{Suzuki1993}.
This is close to optimal, because it is not possible to 
perform simulations in less than linear time, as
Berry et al \cite{Berry2007} prove.
They provide a formula for the optimal
number $k_{\text{opt}}$ of higher order terms to use,
trading off extra operations per step $\tau$
for less steps due to the improved accuracy of each step,
\begin{equation}
k_{\text{opt}} = \left\lceil \frac{1}{2}\sqrt{\log_5(n||\hat{H}||t/\epsilon})\right\rceil
\end{equation}
where $||\hat{H}||$ is the spectral norm of $\hat{H}$ (equal to the magnitude
of the largest eigenvalue for Hermitian matrices).
The corresponding optimal number of operations is bounded by
\begin{equation}
Op_{\text{Berry}} \le 4g^2n^2||\hat{H}||t\exp\left(2\sqrt{\ln 5\ln(n||\hat{H}||t/\epsilon)}\right)
\end{equation}
This is close to linear for large $(n||\hat{H}||t)$.
Recent work by Papageorgiou and Zhang \cite{Papageorgiou2010} 
improves on Berry et al's results, by explicitly incoporating
the dependence on the norms of the largest and next largest of the $\hat{H}_j$
in equation (\ref{eq:Hj}).

Berry et al \cite{Berry2007} also consider more general Hamiltonians,
applicable more to quantum algorithms than quantum simulation.  
For a sparse Hamiltonian, i.e., with no more than a fixed number
of nonzero entries in each column of its matrix representation,
and a black box function which provides one of these entries when queried,
they derive a bound on the number of calls to the black box function
required to simulate the Hamiltonian $\hat{H}$.
When $||\hat{H}||$ is bounded by a constant, the number of calls
to obtain matrix elements scales as
\begin{equation}
O((\log^{*}n)t^{1+1/2k})
\end{equation}
where $n$ is the number of qubits necessary to
store a state from the Hilbert space on which $\hat{H}$ acts, and
$\log^*n$ is the iterative log function, the number of times $\log$
has to be applied until the result is less than or equal to one.
This is a very slowly growing function, for practical values of $n$
it will be less than about five.
This scaling is thus almost optimal, since (as already noted)
sub-linear time scaling is not possible.
These results apply to Hamiltonians where there is no tensor product structure,
so generalise what simulations it is possible to perform efficiently.
Child and Kothari \cite{Childs2009,Childs2010} provide improved
methods for sparse Hamiltonians by decomposing them into sums 
where the graphs corresponding to the non-zero entries are star graphs.
They also prove a variety of cases where efficient simulation of
non-sparse Hamiltonians is possible, using the method developed
by Childs \cite{Childs2008} to simulate Hamiltonians using quantum walks.
These all involve conditions under which an efficient description of
a non-sparse Hamiltonian can be exploited to simulate it.  While of
key importance for the development of quantum algorithms, these
results don't relate directly to simulating physical Hamiltonians.

If we want to simulate bipartite (i.e., two-body) Hamiltonians
$\{\hat{H}^{(2)}_j\}$ using only bipartite Hamiltonians $\{\hat{H}^{(2)}_q\}$,
the control sequences can be efficiently determined
\cite{Wocjan2001a,Wocjan2002,Bennett2002}.
D\"ur, Bremner and Briegel \cite{dur2007} provide detailed prescriptions
for how to map higher-dimensional systems onto pairwise interacting qubits.
They describe three techniques: using commutators between different
$\hat{H}_q$ to build up higher order interactions; graph state encodings;
and teleportation-based methods.  All methods incur a cost in terms of
resources and sources of errors, which they also analyse in detail.
The best choice of technique will depend on the particular problem
and type of quantum computer available.

The complementary problem: given two-qubit Hamiltonians, how can
higher dimensional qubit Hamiltonians be approximated efficiently,
was tackled by Bravyi et al \cite{Bravyi2008}.
They use perturbation theory gadgets to
construct the higher order interactions, which can be viewed as
a reverse process to standard perturbation theory.
The generic problem of $\ell$-local Hamiltonians in an algorithmic setting
is known to be NP-hard for finding the ground state energy, but Bravyi et al
apply extra constraints to restrict the Hamiltonians of both
system and simulation to be physically realistic.
Under these conditions, for many-body qubit Hamiltonians
$\hat{H} = \sum_j\hat{H}^{(\ell)}_j$ with a maximum of $\ell$ interactions per qubit,
and where each qubit appears in only a constant number
of the $\{\hat{H}^{(\ell)}_j\}$ terms,
Bravyi et al show that they can be simulated
using two-body qubit Hamiltonians $\{\hat{H}^{(2)}_q\}$
with an absolute error given by $n \epsilon ||\hat{H}^{(\ell)}_j||_{\text{sup}}$;
where $\epsilon$ is the precision, $||\hat{H}^{(\ell)}_j||_{\text{sup}}$ the
largest norm of the local interactions and $n$ is the number of qubits.
For physical Hamiltonians, the ground state energy
is proportional to $n||\hat{H}^{(\ell)}_j||$, allowing an
efficient approximation of the ground state energy with
arbitrarily small relative error $\epsilon$.

Two-qubit Hamiltonians $\{\hat{H}^{(2)}_q\}$ with local operations
are a natural assumption for modelling a quantum computer, but so
far we have only discussed the interaction Hamiltonian.
Vidal and Cirac \cite{Vidal2002} consider the role of and requirements
for the local operations in more detail, by adding ancillas-mediated
operations to the available set of local operations.
They compare this case with
that of local operations and classical communication (LOCC) only.
For a two-body qubit Hamiltonian, the simulation can 
be done with the same time efficiency, independent of whether ancillas 
are used, and this allows the problem of time optimality to be solved
\cite{Bennett2002}.
However, for other cases using ancillas gives some extra efficiency,
and finding the time optimal sequence of operations is difficult.
Further work on time optimality for the two qubit case
by Hammerer et al \cite{Hammerer2002} and
Haselgrove et al \cite{Haselgrove2003} proves that in most cases,
a time optimal simulation requires an infinite number of infinitesimal
time steps.
Fortunately, they were also able to show that using finite time steps
gives a simulation with very little extra time cost compared to the 
optimal simulation.  
This is all good news for the practical feasibility of useful quantum
simulation.  

The assumption of arbitrary efficient local operations and a fixed
but switchable interaction is not experimentally feasible in all 
proposed architectures.
For example, NMR 
quantum computing has to contend with the extra constraint
that the interaction is always on.  Turning it off when not
required has to be done
by engineering time-reversed evolution using local operations.
The NMR community has thus developed practical solutions to many
Hamiltonian simulation problems of converting one Hamiltonian into
another.  In turn, much of this is based on pulse sequences
originally developed in the 1980s.
While liquid state NMR quantum computation is not scalable, it is an
extremely useful test bed for most quantum computational tasks,
including quantum simulation, and many of the results already mentioned on
universal Hamiltonian simulation owe their development to NMR theory
\cite{Wocjan2001,Bennett2002,Dodd2002}.
Leung \cite{Leung2002} gives
explicit examples of how to do time reversed Hamiltonians for
NMR quantum computation.  Experimental aspects
of NMR quantum simulation are covered in section \ref{NMR}.

The assumption of arbitrary efficient local unitary control operations
also may not be practical for realistic experimental systems.
This is a much bigger restriction than an always on interaction, and
in this case it may only be possible to simulate
a restricted class of Hamiltonians.
We cover some examples in 
the relevant experimental sections.

%% file: Measure.tex
\section{Data extraction}\label{Measure} 

So far, we have discussed in a fairly abstract way
how to evolve a quantum state according to a given Hamiltonian.  
While the time evolution itself is illuminating in a classical
simulation, where the full description of the wavefunction is
available at every time step, quantum simulation gives us only
very limited access to the process.
We therefore have to design our simulation to provide the
information we require efficiently.  The first step is to
manage our expectations: the whole wavefunction is an exponential
amount of information, but for an efficient simulation we can extract
only polynomial-sized results.
Nonetheless, this includes a wide range of properties of quantum
systems that are both useful and interesting, such as
energy gaps \cite{Wu2002};
eigenvalues and eigenvectors \cite{Abrams1999}; and
correlation functions, expectation values and
spectra of Hermitian operators \cite{Somma2002}.
These all use related methods, including phase estimation or
quantum Fourier transforms, to obtain the results.  Brief details
of each are given below in sections \ref{gap} to \ref{corr}.

As will become clear, we may need to use the output of one simulation as the
input to a further simulation, before we can obtain the results we want.
The distinction between input and output is thus somewhat arbitrary,
but since simulation algorithm design is driven by the desired
end result, it makes sense to discuss the most common outputs first.

Of course, many other properties of the quantum simulation can
be extracted using suitable measurements.  Methods developed
for experiments on quantum systems can be adapted for quantum
simulations, such as quantum process tomography 
\cite{OBrien2004} (though this has
severe scaling problems beyond a few qubits), and the more efficient
direct characterisation method of Mohseni and Lidar \cite{Mohseni2006}.
Recent advances in developing polynomially efficient measurement
processes, such as described by Emerson et al \cite{Emerson2007},
are especially relevant.
One well-studied case where a variety of other parameters are
required from the simulation is quantum chaos, described in
section \ref{qchaos}.

\subsection{Energy gaps}\label{gap}

One of the most important properties of an interacting quantum system
is the energy gap between the ground state and first excited state.
To obtain this using quantum simulation,
the system is prepared in an initial state that is a mixture of the ground
and first excited state (see section \ref{adiabatic}).
A time evolution is then performed, which results in a phase difference 
between the two components that is directly proportional to the energy gap.
The standard phase estimation algorithm \cite{Cleve1998},
which uses the quantum Fourier transform,
can then be used to extract this phase difference.
The phase estimation algorithm requires that the simulation
(state preparation, evolution and measurement)
is repeated a polynomial number of times to produce sufficient
data to obtain the phase difference.
An example, where this method is described in detail
for the BCS Hamiltonian, is given by Wu et al \cite{Wu2002}.
The phase difference can also be estimated by measuring
the evolved state using any operator $\hat{M}$
such that $\langle G| \hat{M} |E_1\rangle \ne 0$.
where $|G\rangle$ is the ground state and $|E_1\rangle$ the first excited state.
Usually this will be satisfied for any operator that does not commute
with the Hamiltonian, giving useful experimental flexibility.
A polynomial number of measurements are made, for a range of different times.
The outcomes can then be classically
Fourier transformed to obtain the spectrum, which will have peaks
at both zero and the gap \cite{Brown2006}.  There will be further peaks in
the spectrum if the initial state was not prepared perfectly
and had a proportion of higher states mixed in.
This is not a problem, provided the signal from the gap
frequency can be distinguished, which in turn depends on 
the level of contamination with higher energy states.
However, in the vicinity of a quantum phase transition,
the gap will become exponentially small.  It is then necessary
to estimate the gap for a range of values of the order parameter
either side of the phase transition, to identify when it is
shrinking below the precision of the simulation.
This allows the location of the phase transition to be determined,
up to the simulation precision.

\subsection{Eigenvalues and eigenvectors}\label{eigs}

Generalising from both the Lloyd method for the time evolution
of Hamiltonians and the phase estimation method for finding energy gaps, 
Abrams and Lloyd \cite{Abrams1999} provided an algorithm for
finding (some of) the eigenvalues and eigenvectors of any
Hamiltonian $\hat{H}$ for which $\hat{U}=\exp(i\hat{H}t/\hbar)$
can be efficiently simulated.
Since $\hat{U}$ and $\hat{H}$ share the same 
eigenvalues and eigenvectors, we can equally well use $\hat{U}$
to find them.  Although we can only efficiently 
obtain a polynomial fraction of them, we are generally
only interested in a few, for example the lowest lying energy states.

The Abrams-Lloyd scheme requires an approximate eigenvector
$|V_{a}\rangle$, which must have an overlap $|\langle V_a| V\rangle|^2$
with the actual eigenvector $|V\rangle$ that is not 
exponentially small.  For low energy states, an approximate
adiabatic evolution could be used to prepare a suitable $|V_{a}\rangle$,
see section \ref{adiabatic}.
The algorithm works by using an index register of $m$ qubits
initialised into a superposition of all numbers 0 to $2^m-1$.
The unitary $\hat{U}$ is then conditionally applied to the register
containing $|V_{a}\rangle$ a total of $k$ times, where 
$k$ is the number in the index register.  The components of $|V_{a}\rangle$
in the eigenbasis of $\hat{U}$ now each have a different phase
and are entangled to a different index component.  An inverse
quantum Fourier transform transfers the phases into the index
register which is then measured.  The outcome of the measurement
yields one of the eigenvalues, while the other register now
contains the corresponding eigenvector $|V\rangle$.
Although directly measuring $|V\rangle$ won't yield much useful information,
it can be used as the input to another quantum simulation process
to analyse its properties.

\subsection{Correlation functions and Hermitian operators}\label{corr}

\begin{figure}[tb]
\centering
\includegraphics[width=12cm]{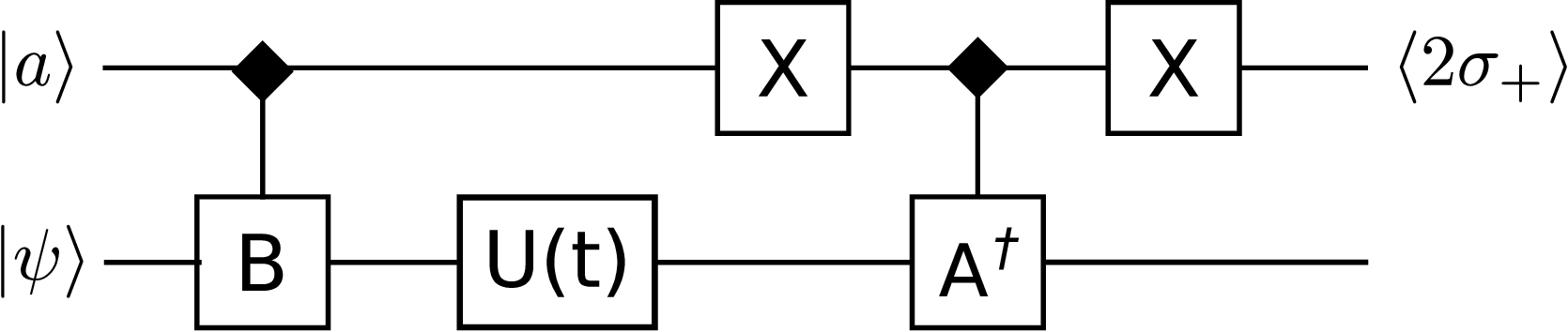}
\caption{A quantum circuit for measuring correlation functions,
X is the Pauli $\sigma_x$ operator,
U(t) is the time evolution of the system, and Hermitian operators $\hat{A}$ 
and $\hat{B}$ are operators (expressible as a sum of unitary operators)
for which the correlation function is required.
The inputs are a single qubit ancilla $|a\rangle$ 
prepared in the state $(|0\rangle + |1\rangle)/\sqrt{2}$
and $|\psi\rangle$, the state of the quantum system for which the
correlation function is required.  $\langle2\sigma_+\rangle$ is
the output obtained when the ancilla is measured in the $2\sigma_+ = 
\sigma_x + \sigma_y$ basis, which provides
an estimate of the correlation function.%
}
\label{Correlation}
\end{figure}

Somma et al \cite{Somma2002} provide detailed methods for
extracting correlation functions, expectation values of
Hermitian operators, and the spectrum of a Hermitian operator.
A similar method is employed for all of these, we describe it
for correlation functions.
A circuit diagram is shown in figure \ref{Correlation}.
This circuit can compute correlation functions of the form 
\begin{equation}
C_{AB}(t) = \langle \hat{U}^{\dagger}(t) \hat{A} \hat{U}(t) \hat{B} \rangle
\end{equation}
where $\hat{U}(t)$ is the time evolution of the system,
and $\hat{A}$ and $\hat{B}$ are expressible as a sum of unitary operators.
The single qubit ancilla $|a\rangle$, initially in the state
$(|0\rangle + |1\rangle)/\sqrt{2}$, is used to control the conditional
application of $\hat{B}$ and $\hat{A}^{\dagger}$, between which the
time evolution $\hat{U}(t)$ is performed.  
Measuring $|a\rangle$ then provides an estimate of the correlation
function to one bit of accuracy.  Repeating the computation 
will build up a more accurate estimate by combining all the outcomes.
By replacing $\hat{U}(t)$ with the space translation operator, spatial
correlations instead of time correlations can be obtained.

\subsection{Quantum chaos}\label{qchaos}

The attractions of quantum simulation caught the imagination of researchers
in quantum chaos relatively early in the development of quantum
computing.
Even systems with only a few degrees of freedom and relatively
simple Hamiltonians can exhibit chaotic behaviour \cite{Georgeot2001b}.
However, classical simulation methods are of limited use for studying
quantum chaos, due to the exponentially growing Hilbert space.
One of the first quantum chaotic systems for which an efficient quantum
simulation scheme was provided is the quantum baker's transformation.
Schack \cite{Schack1998}
demonstrates that it is possible to approximate this map as a sequence
of simple quantum gates using discrete Fourier transforms.
Brun and Schack \cite{Brun1999} then showed
that the quantum baker's map is equivalent to a shift map and
numerially simulated how it would behave on a three qubit NMR
quantum computer.

While the time evolution methods for chaotic dynamics
are straightforward, the important issue is how to
extract useful information from the simulation.
Using the kicked Harper model,
L\'evi and Georgeot \cite{Levi2004}
extended Schack's Fourier transform method to obtain a range of
characteristics of the behaviour in different regimes,
with a polynomial speed up.
Georgeot \cite{Georgeot2004} discusses further methods to extract information
but notes that most give only
a polynomial increase in efficiency over classical algorithms.
Since classical simulations of quantum chaos are generally exponentially
costly, it is disappointing
not to gain exponentially in efficiency in general with a quantum simulation.
However, there are some useful exceptions:
methods for deciding whether a system is chaotic or regular using
only one bit of information have been developed by Poulin et al
\cite{Poulin2003}, and also for measuring the fidelity
decay in an efficient manner \cite{Poulin2004}.
A few other parameters, such as diffusion constants, may also
turn out to be extractable with exponential improvement over classical
simulation.  A review of quantum simulations applied to quantum chaos
is provided by Georgeot \cite{Georgeot2007}.

%% file: Initialise.tex
\section{Initialization} \label{Initialisation}

As we saw in the previous section, a crucial step in 
extracting useful results from a quantum simulation is
starting from the right initial state.
These will often be complex or unknown states,
such as ground states and Gibbs thermal states.
Preparing the initial state is thus as important as the time
evolution, and significant research has gone into 
providing efficient methods.
An arbitrary initial state takes exponentially many
parameters to specify, see equation (\ref{eq:qstate}),
and hence exponential time to prepare using its description.
We can thus only use states which have
more efficient preparation procedures.  
Although preparing an unknown state sounds like it should be even
harder than preparing a specific arbitrary state, when a
simple property defining it is specified,
there can be efficient methods to do this.

\subsection{Direct state construction}\label{direct}

Where an explicit description is given for the initial state
we require, it can be prepared using any method for
preparing states for a quantum register.  
Soklakov and Schack \cite{Soklakov2004,Soklakov2005} provide a method
using Grover's search algorithm, that is efficient provided
the description of the state is suitably efficient.
Plesch and Brukner \cite{Plesch2010} optimise general
state preparation techniques to reduce the prefactor in
the required number of CNOT gates to close to the optimal
value of one.  Direct state preparation is thus feasible for
any efficiently and completely specified pure initial state.
Poulin and Wocjan \cite{Poulin2008} analyse the efficiency of
finding ground states with a quantum computer.  This is
known to be a QMA-complete problem for $k$-local Hamiltonians
(which have the form of equation (\ref{eq:Hj}) where the $H_j$ involve
$k$ of the variables, for $k\ge 2$).
They provide a method based on Grover's search, with some sophisticated
error reduction techniques, that gives a quadratic speed up
over the best classical methods for finding eigenvalues of matrices.
Their method is really a proof of the complexity of the problem
in general rather than a practical method for
particular cases of interest, which may not be as hard as the
general case they treat.

\subsection{Adiabatic evolution}\label{adiabatic}

Adiabatic quantum computing encodes the problem into the ground state
of a quantum Hamiltonian.
The computation takes place by evolving the Hamiltonian from one with an easy
to prepare ground state $\hat{H}_0$ to the one with the desired solution 
$\hat{H}_1$ as the ground state,
\begin{equation}
\hat{H}_{\text{ad}} = (1-s(t))\hat{H}_0 + s(t)\hat{H}_1
\end{equation}
where the monotonically increasing function $s(t)$  controls the rate of
change, $s(0) = 0$.
This has to be done slowly enough, to keep the system in the ground state
throughout.
Provided the gap between the ground state and first excited state
does not become exponentially small, ``slowly enough'' will require
only polynomial time.
Extensive discussion of quantum adiabatic state preparation from
an algorithmic perspective, including other useful states that
can be produced by this method, is given by Aharonov and Ta Shma
\cite{Aharonov2003}.

The application to preparing ground states for quantum
simulation was first suggested by Ortiz et al \cite{Ortiz2001}.
The potential issue is that finding
ground states is in general a QMA-complete problem, which implies
it may not be possible to do this efficiently for all cases of interest,
that the gap will become exponentially small at some point in the evolution.
In particular, we know the gap will become exponentially small
if the evolution passes through a quantum phase transition.
Since the study of quantum phase transitions is one aspect of 
quantum many-body systems of interest for quantum simulation,
this is not an academic problem, rather, it is likely to occur in practice.
Being of crucial importance for adiabatic quantum computation,
the question of how the time evolution scales near a
phase transition has been extensively studied.  Recent work
by Dziarmaga and Rams \cite{dziarmaga10a} on inhomogeneous quantum
phase transitions explains how in many cases of practical interest,
disruption to the adiabatic evolution across the phase transition
can be avoided.  An inhomogeneous phase transition is where the
order parameter varies across the system.  Experimentally, this
is very likely to happen to some extent, due to the difficulty of
controlling the driving mechanism perfectly, the strength of
a magnetic field for example.
Consequently, the phase change will also happen at slightly different
times for different parts of the system, and there will be boundaries
between the different regions.  Instead of being a global change,
the phase transition sweeps through the system, and the speed with
which the boundary between the phases moves can be estimated.
Provided this is slower than the timescale on which local transitions
take place, this allows the region in the new phase to influence the
transition of the nearby regions.  The end result is that it 
is possible to traverse the phase transition in polynomial time
without ending up in an excited state, for a finite-sized system.  

Moreover, we don't generally need to prepare a pure ground state for
quantum simulation of such systems.
The quantity we usually
wish to estimate for a system with an unknown ground state is
the energy gap between the ground and first excited states.
As described in section \ref{gap},
this can be done by using phase estimation applied to a
coherent superposition of the ground state and first excited state.
So traversing the adiabatic evolution only approximately, to
allow a small probability of exciting the system is in fact
a useful state preparation method.
And if we want to obtain the lowest eigenvalues and study the 
corresponding eigenvectors of a Hamiltonian, again we only need
a state with a significant proportion of the ground state as
one component, see section \ref{eigs}.

Oh \cite{Oh2008} describes a refinement of the Abrams-Lloyd
method for finding eigenvalues and eigenvectors described in
section \ref{eigs}, in which the state preparation using
the adiabatic method is run in parallel with the phase estimation
algorithm for estimating the ground state energy.
This allows the ground state energy to be extracted
as a function of the coupling strength that is increased
as the adiabatic evolution proceeds.
Oh adds an extra constant energy term to the Hamiltonian, to
tune the running time, and uses the Hellman-Feynman theorem to
obtain the expectation value of the ground state observable.
Boixo et al \cite{Boixo2009} prove that this and related methods
using continuous measurement, as provided by the
phase estimation algorithm run in parallel,
improve the adiabatic state preparation.
The running time is inversely proportional to the
spectral gap, so will only be efficient when the gap
remains sufficiently large throughout the evolution.

\subsection{Preparing thermal equilibrium states}\label{thermal}

Temperature dependent properties of matter are of key importance.
To study these, efficiently preparing thermal states for quantum
simulation is crucial.
The most obvious method to use is to actually equilibriate the
quantum state to the required temperature, using a heat bath.
Terhal and DiVincenzo \cite{Terhal2000} describe how this can
be done with only a relatively small bath system,
by periodically reinitializing the bath to the required temperature.
The core of this algorithm begins by initializing the
system in the ``all zero'' state, $|00\dots 00\rangle\langle00\dots 00|$
and the bath in an equilibrium state of the required temperature.
The system and bath are then evolved for time $t$ after which the bath
is discarded and re-prepared in its equilibrium state.
This last step is repeated a number of times, creating
to a good approximation the desired thermal initial state for subsequent
simulation.
Terhal and DiVincenzo don't give explicit bounds on the running
time of their method, though they do discuss reasons why they don't 
expect it to be efficient in the general case.
Recent results from Poulin and Wocjan \cite{Poulin2009}
prove the upper bound on the running time for thermalisation is $D^a$, where $a\le 1/2$
is related to the Helmholtz free energy of the system, and $D$ is
the Hilbert space dimension.  This thus confirms
that Terhal and DiVincenzo's method may not be efficient in general.
Poulin and Wocjan also provide a method for approximating the
partition function of a system with a running time proportional
to the thermalisation time.  The partition function is
useful because all other thermodynamic quantities of interest
can be derived from it.  So in cases where their method
can be performed efficiently, it may be prefered over
the newly developed quantum Metropolis algorithm described next.

The quantum Metropolis algorithm of Temme et al \cite{Temme2009} 
is a method for efficiently sampling from any
Gibbs distribution. It is the quantum
analogue of the classical Metropolis method.
The process starts from 
a random energy eigenstate $|\Psi_i\rangle$ of energy $E_i$.
This can be prepared efficiently by evolving from any initial state
with the Hamiltonian $\hat{H}$, then using
phase estimation to measure the energy and thereby project into an eigenstate.
The next step is to generate
a new ``nearby'' energy eigenstate $|\Psi_j\rangle$ of energy $E_j$.
This can be achieved via a local random unitary
transformation such that
$|\Psi_i\rangle \longrightarrow \sum_j c_{ij} |\Psi_j\rangle$ with $E_j \sim E_i$.
Phase estimation is then used again to project into the state
$|\Psi_j\rangle$ and gives us $E_j$.
We now need to accept the new configuration with probability
$p_{ij} = \min[1, \exp(-\beta(E_i - E_j))]$, where $\beta$ is inverse temperature.
Accepting is no problem, the state of the quantum registers are
in the new energey eigenstate $|\Psi_j\rangle$ as required.
The key development in this method is how to reject, which requires
returning to the previous state $|\Psi_i\rangle$.
By making a very limited measurement that determines only
one bit of information (accept/reject), the coherent part of the
phase estimation step can be reversed with high probability; 
repeated application of the reversal steps can increase the
probability as close to unity as required.  Intermediate
measurements in the process indicate when the reversal has
succeeded, and the iteration can be terminated.
The process is then repeated to obtain
the next random energy state in the sequence.
This efficiently samples from the thermal distribution for preparing
the initial state, and can be used for any type of quantum system,
including fermions and bosons.
Temme et al also prove that their algorithm correctly 
samples from degenerate subspaces efficiently.

%% file: Evolve.tex
\section{Hamiltonian evolution}\label{evolve}

The Lloyd method of evolving the quantum state in time
according to a given Hamiltonian, described in section \ref{Lloyd},
is a simple form of numerical integration.
There are a variety of other methods for 
time evolution of the dynamics in classical simulation,
some of which have been adapted for quantum simulation.
Like their classical counterparts, they provide 
significant advantages for particular types of problem.
We describe two of these methods that are especially promising
for quantum simulation: a quantum version of
the pseudo-spectral method using quantum Fourier transforms, and
quantum lattice gas automata.  Quantum chemistry has also
developed a set of specialised simulation methods for
which we describe some promising quantum counterparts in section \ref{qchem}.
We would also like to be able to simulate systems subject to noise
or disturbance from an environment, open quantum systems.  Some
methods for efficiently treating non-unitary evolution are
described in section \ref{open}.

\subsection{Quantum pseudo-spectral method} \label{Fourier}

Fast Fourier transforms are employed extensively in classical computational
methods, despite incurring a significant computational cost.  Their 
use can simplify the calculation in a wide diversity of
applications.  When employed for dynamical evolution, the
pseudo-spectral method converts between real space and Fourier space
(position and momentum) representations.  This allows terms to be
evaluated in the most convenient representation, providing 
improvements in both the speed and accuracy of the simulation.

The same motivations and advantages apply to quantum simulation.
A quantum Fourier transform can be implemented efficiently on
a quantum computer for any quantum state \cite{Jozsa1998,Browne2006}.
Particles moving in external potentials often have
Hamiltonians with terms that are diagonal in 
the position basis plus terms that are diagonal in the momentum basis.
Evaluating these terms in their diagonal bases 
provides a major simplification to the computation.
Wiesner \cite{Wiesner1996} and Zalka \cite{Zalka1998a, Zalka1998}
gave the first detailed descriptions of this approach for particles
moving in one spatial dimension, and showed that
it can easily be generalized to a many particle 
Schr\"odinger equation in three dimensions.
To illustrate this, consider the one-dimensional Schr\"odinger equation
(with $\hbar=1$),
\begin{equation}
i\frac{\partial}{\partial t} \Psi(x,t) = 
\left(-\frac{1}{2m}\nabla^2 + V(x)\right)\Psi(x,t)
\end{equation}
for a particle in a potential $V(x)$.  As would be done for a classical
simulation this is first discretized so the position is approximated on
a line of $N$ positions (with periodic boundary conditions)
and spacing $\Delta x$.  We can then write the wavefunction as
\begin{equation}
|\Psi(n,t)\rangle = \sum_n a_n(t)|n\rangle
\end{equation}
where $\{|n\rangle\}$, $0\le n<N$ are position basis states, and $a_n(t)$
is the amplitude to be at position $n$ at time $t$.
For small time steps $\Delta t$, the Green's function to evolve
from $x_1$ to $x_2$ in time $\Delta t$ becomes 
\begin{equation}
G(x_1,x_2,\Delta t) = \kappa\exp\left\{i\frac{m}{2}\frac{(x_1-x_2)^2}{\Delta t}
+iV(x_1)\Delta t\right\}
\end{equation}
where $\kappa$ is determined by the normalization.
The transformation in terms of basis states is the inverse of this,
\begin{equation}
G'(n,n',\Delta t)|n\rangle = \frac{1}{\sqrt{N}}\sum_s{n'}
\exp\left\{-i\frac{m}{2}\frac{(n-n')^2\Delta x^2}{\Delta t} -iV(n\Delta x)\Delta t\right\}
|n'\rangle
\end{equation}
Expanding the square, this becomes
\begin{eqnarray}
G'(n,n',\Delta t)|n\rangle &=& 
\frac{1}{\sqrt{N}}\exp\left\{-i\frac{m}{2}\frac{n^2\Delta x^2}{\Delta t} -iV(n\Delta x)\Delta t\right\}\nonumber\\
&\times&\sum_{n'}\exp\left\{-im\frac{nn'\Delta x^2}{\Delta t}\right\}\left(
\exp\left\{-i\frac{m}{2}\frac{n'^2\Delta x^2}{\Delta t} \right\}\right)
|n'\rangle
\label{eq:psm}
\end{eqnarray}
The form of equation \ref{eq:psm} is now two diagonal matrices with a Fourier
transform between them, showing how the pseudo-spectral method
arises naturally from standard solution methods.
Benenti and Strini \cite{Benenti2007} provide a 
pedagogical description of this method applied to a 
single particle, with quantitative analysis
of the number of elementary operations required for small
simulations.  They estimate that, for present day capabilities of
six to ten qubits, the number of operations required for a
useful simulation is in the tens of thousands, which is many more than
can currently be performed coherently.
Nonetheless, the efficiency savings over the Lloyd method
will still make this the preferred option whenever the terms in the
Hamiltonian are diagonal in convenient bases related by a
Fourier transform.

\subsection{Lattice gas automata}\label{LGA}

Lattice gas automata and lattice-Boltzmann methods
are widely used in classical simulation
because they evolve using only local interactions,
so can be adapted for efficient parallel processing.
Despite sounding like abstract models of physical systems,
these methods are best understood as sophisticated techniques to
solve differential equations: the ``gas'' particles have nothing 
directly to do with the particles in the system they are simulating.
Instead, the lattice gas dynamics are shown to correspond to the
differential equation being studied in the continuum limit of
the lattice.  Different equations are obtained from different
local lattice dynamics and lattice types.  Typically, a
face-centred cubic or body-centred cubic lattice is required,
to ensure mixing of the particle momentum in different directions
\cite{Frisch1986}.
Succi and Benzi \cite{Succi1993} developed a lattice Boltzmann
method for classical simulation of quantum systems, and Meyer
\cite{Meyer1996} applied lattice gas automata to many-particle Dirac systems.
Boghosian and Taylor \cite{Boghosian1998a} built on this work to
develop a fully quantum version of lattice gas automata, 
and showed that this can be efficiently implemented on a qubit-based quantum
computer, for simulations of many interacting quantum particles
in external potentials.  
This method can also be applied to the many-body Dirac equation (relativistic 
fermions) and gauge field theories, by suitably modifying the lattice
gas dynamics, both are briefly discussed by Boghosian and Taylor.

To illustrate the concept, we describe a simple quantum lattice gas in
one dimension.  This can be encoded into two qubits
per lattice site, one for the plus direction and the other for the minus
direction.  The states of the qubits represent
$|1\rangle$ for a particle present, and $|0\rangle$ for no particle,
with any superposition between these allowed.  Each time step
consists of two operations, a ``collision'' operator that
interacts the qubits at each lattice site, and a ``propagation''
operator that swaps the qubit states between neighboring lattice sites,
according to the direction they represent.  This is like a coined
quantum walk dynamics, which is in fact a special case of lattice gas
automata, and was shown to correspond to the Dirac equation
in the continuum limit by Meyer \cite{Meyer1996}.
Following Boghosian and Taylor \cite{Boghosian1998},
we take the time step operator $S.C$ combining both collision $C$
and propagation $S$ to be
\begin{equation}
S.C \left(\begin{array}{c}q_1(x+1,t+1)\\ q_2(x-1,t+1)\end{array}\right) =
\frac{1}{2}\left( \begin{array}{cc}
        1-i & -1-i\\
        -1-i & 1+i
        \end{array} \right)\left(\begin{array}{c} q_1(x,t)\\q_2(x,t)\end{array} \right)
\end{equation}
where $q_1$ and $q_2$ are the states of the two qubits.
Taking the continuum limit where the lattice spacing scales as
$\epsilon$ while the time scales as $\epsilon^2$ gives
\begin{equation}
\frac{\partial}{\partial t}q_1(x,t) = \frac{i}{2}\frac{\partial^2}{\partial x^2}q_2(x,t)
\end{equation}
and a similar equation interchanging $q_1$ and $q_2$.  Hence for the sum,
\begin{equation}
\frac{\partial}{\partial t}\{q_1(x,t)+q_2(x,t)\} = \frac{i}{2}\frac{\partial^2}{\partial x^2}\{q_1(x,t) + q_2(x,t)\}
\end{equation}
The total amplitude $\psi(x,t) = q_1(x,t) + q_2(x,t)$ thus satisfies a
Schr\"odinger equation.  It is a straightforward generalisation to
extend to higher dimensions and more particles, and to add interactions
between particles and external potentials, as explained in detail
by Boghosian and Taylor \cite{Boghosian1998}.
Based on the utility of lattice gas automata methods for classical simulation,
we can expect these corresponding quantum versions to prove highly
practical and useful when sufficiently large quantum computers
become available.

\subsection{Quantum chemistry}\label{qchem}

Study of the dynamics and properties of molecular reactions
is of basic interest in chemistry and related areas.  Quantum effects
are important at the level of molecular reactions, but exact
calculations based on a full Schr\"odinger equation for all the
electrons involved are beyond the capabilities of classical computation,
except for the smallest molecules.  A hierarchy of approximation
methods have been developed, but more accurate calculations would
be very useful.  
Aspuru-Guzik et al \cite{Aspuru-Guzik2005} study the application of
quantum simulation to calculation of the energies of small
molecules, demonstrating that a quantum computer can obtain the 
energies to a degree of precision greater than that required by 
chemists for understanding reaction dynamics,
and better than standard classical methods.
To do this, they adapt the method of Abrams and Lloyd \cite{Abrams1999}
for finding the eigenvalues of a Hamiltonian described in
section \ref{eigs}.
The mapping of the description of the molecule to qubits is
discussed in detail, to obtain an efficient representation.  
In a direct mapping, the qubits are used to store the occupation
numbers for the atomic orbitals: the Fock space of the molecule is 
mapped directly to the Hilbert space of the qubits.
This can be compacted by restricting to the subspace
of occupied orbitals, i.e., fixed particle number,
and a further reduction in the number of qubits required 
is obtained by fixing the spin states of the electrons as well.
By doing classical simulations of the quantum simulation for
H$_{2}$O and LiH, they show in detail that these methods are
feasible.  For the simulations, they use the Hartree-Fock 
approximation for the initial ground state.  In some situations, however, this 
state has a vanishing overlap with the actual ground state. This means 
it may not be suitable in the dissociation limit or in the 
limit of large systems.  A more accurate approximation of the required
ground state can be prepared using adiabatic evolution, see section
\ref{adiabatic}.  Aspuru-Guzik et al confirm numerically that this
works efficiently for molecular hydrogen.  Data from experiments or
classical simulations can be used
to provide a good estimate of the gap during the adiabatic evolution,
and hence optimise the rate of transformation between the initial and
final Hamiltonians.

The Hartree-Fock wavefunctions used by Aspuru-Guzik et al
are not suitable for excited states.
Wang et al \cite{Wang2008} propose using an initial 
state that is based on a multi-configurational self consistent field (MCSCF).
These initial states are also suitable for strong 
interactions, since they avoid convergence to unphysical states when the 
energy gap is small. In general, using MCSCF wavefunctions allows 
an evolution that is faster and safer than using Hartree-Fock
wavefunctions, so represents a significant improvement. 

To calculate the properties of chemical reactions classically,
the Born-Oppenheimer approximation is used for the electron
dynamics.
The same can be done for quantum simulations; however,
Kassal et al \cite{Kassal2008} observe that,
for all systems of more than four atoms, 
performing the exact computation on a quantum computer 
should be more efficient.
They provide a detailed method for exact simulation of
atomic and molecular electronic wavefunctions, based on 
discretizing the position in space,
and evolving the wavefunction using the QFT-based time 
evolution technique presented by Wiesner \cite{Wiesner1996} and Zalka 
\cite{Zalka1998} described in section \ref{Fourier}. 
Kassal et al discuss three approaches
to simulating the interaction potentials and provide
the initialisation procedures needed for each, along with
techniques for determining reaction probabilities,
rate constants and state-to-state transition probabilities.
These promising results suggest that quantum chemistry will feature
prominently in future applications of quantum simulation.

\subsection{Open quantum systems}\label{open}

Most real physical systems are subject to noise from their
environment, so it is important to be able to include this 
in quantum simulations.  For many types of environmental decoherence,
this can be done as a straightforward extension to Lloyd's basic
simulation method  \cite{Lloyd1996} 
(described in section \ref{Lloyd}).
Lloyd discusses how to
incorporate the most common types of environmental decoherence
into the simulation.
For uncorrelated noise, the appropriate superoperators can be used in place 
of the unitary operators in the time evolution, because these will also 
be local.  Even for the worst case of correlated noise, the environment can 
be modeled by doubling the number of qubits and employing local 
Hamiltonians for the evolution of the environment and its coupling, as 
well as for the system.
Techniques for the simulation of open quantum systems for a single
qubit have been further refined and 
developed by Bacon et al \cite{Bacon2001},
who provide a universal set of processes to simulate
general Markovian dynamics on a single qubit.
However, it is not known whether these results can be extended to include
all Markovian dynamics in systems
of more than two qubits, since it is no longer possible to write the dynamics 
in the same form as for the one and two qubit cases.


Better still, from the point of view of efficiency, is if the effects
of noise can be included simply by using the inevitable decoherence
on the quantum computer itself.
This will work provided the type of decoherence is sufficiently
similar in both statistics and strength.
Even where the aim is to simulate perfect unitary dynamics,
small levels of imperfection due to noisy gates in the simulation may 
still be tolerable, though the unfavorable scaling of
precision with system size discussed in section \ref{noise}
will limit this to short simulations.
Nonetheless, in contrast to the error correction necessary
for digital quantum computations where precise numerical
answers are required, a somewhat imperfect quantum simulation
may be adequate to provide us with a near perfect simulation
of an open quantum system.

%% file: FB.tex
\section{Fermions and bosons}\label{Fermi}

Simulation of many-body systems of interacting fermions
are among the most difficult to handle with classical methods,
because the change of sign when two identical fermions are
exchanged prevents the convergence of classical statistical
methods, such as Monte Carlo sampling.  This is known as the ``sign problem'',
and has limited effective simulation of fermionic many-body systems 
to small sizes that can be treated without these approximations. 
Very recent work from Verstraete and Cirac \cite{verstraete10a} has
opened up variational methods for fermionic systems,
including relativistic field theories \cite{haegeman10a}.
Nonetheless, the computational cost of accurate classical simulations is
still high, and we have from Ortiz et al \cite{Ortiz2001} a general proof
that conducting a simulation of a fermionic system on a quantum computer
can be done efficiently and does not suffer from the sign problem.
They also confirm that errors within the quantum computation
don't open a back door to the sign problem. 
This clears the way for developing detailed algorithms
for specific models of fermionic systems of particular interest.
Some of the most important open questions a quantum computer
of modest size could solve are models involving strongly interacting
fermions, such as for high temperature superconductors.

\subsection{Hubbard model}\label{hubbard}

One of the important fermionic models that has received detailed
analysis is the Hubbard model, one of the most basic microscopic descriptions
of the behaviour of electrons in solids.  Analytic solutions are
challenging, especially beyond one dimension, and while ferromagnetism
is obtained for the right parameter ranges, it is not known whether
the basic Hubbard model produces superconductivity.  The difficulties of
classical simulations thus provide strong motivation
for applying quantum simulation to the Hubbard model.
The  Hubbard Hamiltonian $\hat{H}_{\gamma V}$ is 
\begin{equation}
\hat{H}_{\gamma V} = -\gamma \sum_{\langle j,k \rangle, \sigma} \hat{C}^{\dagger}_{j,\sigma} \hat{C}_{k,\sigma} + V\sum_{j}\hat{n}_{j,\uparrow}\hat{n}_{j,\downarrow}
\label{Hubbard}
\end{equation}
where $ \hat{C}^{\dagger}_{j,\sigma} \hat{C}_{k,\sigma}$ are the fermionic
creation and annihilation operators, $\sigma$ is the spin (up or down),
$\hat{n}_{j,\uparrow}\hat{n}_{j,\downarrow}$
are the number operators for up spin and down spin
states at each site $j$, $\gamma$ is the strength of 
the hopping between sites, and $V$ is the on site potential.
Abrams and Lloyd \cite{Abrams1997} describe two different encodings
of the system into the quantum simulator.
An encoding using the second 
quantization is more natural since the first quantization encoding 
requires the antisymmetrization of the wavefunction ``by hand''.
However, when the 
number of particles being simulated is a lot lower than the available 
number of qubits, the first quantization is more efficient.
In second quantization, there are four possible states each site can be
in: empty, one spin up, one spin down, and a pair of opposite spin.
Two qubits per site are thus required to encode which of the four
states each site is in.  It is then a simple extension of the Lloyd
method to evolve the state of the system according to the
Hubbard Hamiltonian.
Somma et al \cite{Somma2002} describe how to use this method to
find the energy spectrum of the Hubbard 
Hamiltonian for a fermionic lattice system. They perform a classical computer 
simulation of a quantum computer doing a quantum simulation,
to demonstrate the feasibility of the quantum simulation.
The Hubbard model is the natural Hamiltonian in optical
lattice schemes, so there has been considerable
development towards special purpose simulators based on
atoms in optical lattices, these are discussed in section \ref{atomOL}.

\subsection{The BCS Hamiltonian}\label{BCS}

Pairing Hamiltonians are an important 
class of models for many-body systems in which pairwise interactions
are typically described using fermionic (or bosonic) creation and
annihilation operators $\{c_{m}, c^{\dagger}_{m}\}$.
Nucleons in larger atomic nuclei can be described by pairing Hamiltonians,
and
Bardeen, Copper and Schrieffer (BCS) \cite{BCS57}
formulated a model of superconductivity as a a pairing Hamiltonian
in the 1950s.
The BCS model of superconductivity is still not fully understood,
so quantum simulations could be useful to improve our knowledge of
superconducting systems, especially for realistic materials with 
imperfections and boundary effects.
While the BCS ansatz is exact in the thermodynamic limit,
it is not known how well it applies to small systems \cite{knill00a}.

The BCS Hamiltonian for a fully general system can be written
\begin{equation}
H_{BCS}=\sum^{N}_{m=1} \frac{\epsilon_{m}}{2}(c^{\dagger}_{m}c_{m} + c^{\dagger}_{-m}c_{-m}) \,+ \sum^{N}_{m,l=1}V_{ml}c^{\dagger}_{m}c^{\dagger}_{-m}c_{-l}c_{l}
\label{eq:BCS}
\end{equation}
where the parameters $\epsilon_{m}$ and $V_{ml}$ specify the self energy of
the $m$th mode and the interaction energy of the $m$th and $l$th modes
respectively, while $N$ is the total number of occupied modes (pairs of
fermions with opposite spin).
Wu et al \cite{Wu2002} developed a detailed method for
quantum simulation of equation (\ref{eq:BCS}).
The two terms in the BCS Hamiltonian do not commute, therefore the
simulation method requires the use of Trotterization
(see section \ref{Lloyd}) so the two
parts can be individually applied alternately.
This means that any simulation on a universal quantum
computer will require many operations to step through
the time evolution, which will stretch the
experimentally available coherence times.
Savings in the number of operations are thus
important, and recent work by Brown at al \cite{brown10b} adapting
the method to a qubus architecture reduces the number of operations
required in the general case from $O(N^5)$ for NMR to $O(N^2)$ for the qubus.
Pairing Hamiltonians are used to describe many processes in condensed matter
physics and therefore a technique for simulating the BCS Hamiltonian
should be adaptable to numerous other purposes.

\subsection{Initial state preparation}\label{fbstates}

For simulation on qubit quantum computers (as opposed to special
purpose quantum simulators), we first need an efficient
mapping between the particles being simulated and the spin-1/2 algebra
of the qubit systems. Somma et al \cite{Somma2003}
discuss in detail how to map physical particles onto spin-1/2 systems.
For fermions there
is a one-to-one mapping between the fermionic and spin-1/2 algebras.
Particularly in the second quantization this allows a simple mapping that
can be generalised to all anyonic systems which obey the Pauli exclusion
principle, or generalised versions of it.
For bosonic systems there is no direct mapping between the
bosonic algebra and spin 1/2 algebra. Therefore Somma et al
propose using a direct mapping between the state of the
two systems, provided there is a limit on the number of bosons per state.
This mapping is less efficient but allows simulations to be conducted on
the bosonic systems.

Systems of indistinguishable particles require special state preparation
to ensure the resulting states have the correct symmetry.
Ortiz et al \cite{Ortiz2001} developed a method for fermions that was then
adapted for bosons by Somma et al \cite{Somma2003}.
In general, a quantum system of $N_{e}$ fermions 
with an anti-symmetrized wavefunction $|\Psi_e\rangle$
can be written as a sum of Slater determinants $|\Phi_{\alpha}\rangle$
\begin{equation}
|\Psi_e\rangle = \sum^{n}_{\alpha=1} \mathbf{a}_{\alpha}|\Phi_{\alpha}\rangle
\end{equation}
where $n$ is an integer and $\sum_{\alpha=1}^{n}| 
\mathbf{a}_{\alpha}|^2 = 1$.
The individual Slater determinants can be prepared efficiently using 
unitary operations.
Provided the desired state doesn't require an exponential sum of
Slater determinants, with the help of $n$ ancilla qubits
it is possible to prepare the state
\begin{equation}
\sum^{n}_{\alpha=1}\mathbf{a}_{\alpha}|\alpha\rangle\otimes | \Phi_{\alpha}\rangle
\end{equation}
where $|\alpha\rangle$ is a state of the ancilla with the $\alpha$'th
qubit in state $|1\rangle$ and the rest in state $|0\rangle$.
A further register of $n$ ancillas is then used to convert the
state so that there is a component with the original ancillas
in the all zero state,
\begin{equation}
\sum^{n}_{\alpha=1} \mathbf{a}_{\alpha}|0\rangle_{a}\otimes|\Phi_{\alpha}\rangle
\end{equation}
associated with the required state of the fermions.
A measurement in the $z$-basis selects this outcome (all zeros)
with a probability of $1/n$.
This means the preparation should be possible using an average of $n$ trials.

A general bosonic system can be written as a linear combination of 
product states. These product states can be mapped onto spin states and 
then easily prepared by flipping the relevant spins. Once the bosonic 
system has been written as a linear combination of these states, a very 
similar preparation procedure to the one for fermionic systems can be used
\cite{Somma2003}.

While the above method using Slater determinants is practical when
working in second quantization, this isn't always convenient for
atomic and molecular systems.  
Ward et al \cite{Ward2009} present a system for efficiently converting states
prepared using Slater determinants in second quantization 
to a first quantization representation on a real space lattice.
This can be used for both pure and mixed states.

\subsection{Lattice gauge theories}

Lattice gauge theories are important in many areas of physics, and
one of the most important examples from a computational perspective is
quantum chromodynamics (QCD).
Classical QCD simulations are extremely computationally intensive,
but very important for predicting the properties of fundamental
particles.  Providing more efficient quantum simulations would be very
useful to advance the field.
The quantum lattice gas
method developed by Boghosian and Taylor \cite{Boghosian1998}
(discussed in section \ref{LGA}) is suitable for simulating
lattice gauge theories, using similar methods to the lattice QCD
simulations currently performed classically, but with the
benefit of a quantum speed up.
Byrnes and Yamamoto \cite{Byrnes2006} provide a more general
method.
They map the desired Hamiltonian to one involving only Pauli operations
and one and two qubit interactions.
This is then suitable for any qubit-based universal quantum simulator.
They focus on the U(1), SU(2) and SU(3) models,
but their methods easily generalise to higher order SU(N) theories.
To conduct the simulation efficiently it is necessary to use
a truncated version of the model, to keep the number of qubits finite.
They demonstrate that the number of operations required for the time
evolution and for the preparation of the necessary initial states
are both efficient.
To get results inaccessible to classical computers, of the
order of $10^{5}$ qubits will be required. Despite this, the algorithm
has advantages over classical techniques because the calculations are
exact up to a cut off, and with simple adaptions it can be extended to
to simulate fermionic systems.

Methods suitable for special purpose quantum simulators have
been presented by Sch\"utzhold and Mostame \cite{Schutzhold2005}
and Tewari et al \cite{Tewari2006}.
Sch\"utzhold and Mostame describe how to simulate
the O(3) nonlinear $\sigma$-model,
which is of interest to the condensed matter
physics community as it applies to spin systems.
It also reproduces many of the key properties of QCD,
although it is only a toy model in this context.
To conduct their simulation, Sch\"utzhold and Mostame
propose using hollow spheres to trap electrons, described
in more detail in section \ref{electrons}.
Tewari et al \cite{Tewari2006} focus specifically on compact U(1)
lattice gauge theories that are appropriate for dipolar bosons in
optical lattices.  The basic Hamiltonian in optical lattices is
the Hubbard Hamiltonian, equation (\ref{Hubbard}), but different
choices of atom can enhance the Hamiltonian with different
nearest neighbour interactions.  The specific example
chosen by Tewari et al is chromium, which has a magnetic
dipolar interaction that can provide the extra term in the Hamiltonian.
The ratio of the two types of couplings (Hubbard and dipolar) can be
varied over a wide range by tuning the Hubbard interaction strength
using Feshbach resonances.  Further types of relativistic
quantum field theories that can be simulated by atoms in optical lattices
are presented by Cirac et al \cite{Cirac2010}.  

This concludes the theory part of our review, and provides a natural
point to move over to consideration of the different physical
architectures most suited to quantum simulation.

%% file: Over.tex
\section{Overview}\label{over}

As we have seen, 
while algorithms for quantum simulation are interesting in their own 
right, the real drive is towards actual implementations of a useful
size to apply to problems we cannot solve with classical computers.
The theoretical studies show that quantum simulation can be
done with a wide variety of methods and systems, giving plenty of
choices for experimentalists to work with.  Questions remain
around the viability of longer simulations, where errors may
threaten the accuracy of the results, and long sequences of
operations run beyond the available coherence times.
As with quantum computing in general, the main challenge
for scaling up is controlling the
decoherence and correcting the errors that accumulate from
imperfect control operations.  Detailed treatment of these issues
is beyond the scope of this review and well-covered elsewhere
(see, for example, Devitt et al \cite{Devitt2009}).
The extra concern for quantum simulation lies in the
unfavorable scaling of errors with system size, as
discussed in section \ref{noise}.

In section \ref{Universal} we described how to obtain universal
quantum simulation from particular sets of resources, mainly
a fixed interaction with local unitary controls.  Building a universal
quantum simulator will allow us to efficiently simulate any quantum
system that has a local or efficiently describable Hamiltonian.
On the other hand, the generality
of universal simulation may not be necessary if the problem we are
trying to solve has a specific Hamiltonian with properties or
symmetries we can exploit to simplify the simulation.
If the Hamiltonian we want to simulate can be matched with
a compatible interaction Hamiltonian in the quantum simulator,
then there are are likely to be further efficiencies available
through simpler control sequences for converting one into the other.
From the implementation perspective, a special purpose simulator
may be easier to build and operate, a big attraction in these
early stages of development.
Most architectures for quantum computing are also suitable
for universal quantum simulation.  
However, the range of experimental possibilities is broader
if we are willing to specialise to the specific Hamiltonians in the
quantum simulator.  This allows more to be achieved with the same hardware,
and is thus the most promising approach for the first useful
quantum simulations.

Buluta and Nori \cite{Buluta2009} give a brief
overview of quantum simulation that focuses
on the various possible architectures
and what sort of algorithms these could be used for.
There is broad overlap of relevant experimental techniques
with those for developing quantum computers in general,
and many issues and considerations are common to all
applications of quantum computers.
In this paper, we concentrate on implementations that
correspond to the theoretical aspects we have covered. 
Many experimental implementations of quantum simulation to date
have been in NMR quantum computers.  This is not a scalable
architecture, but as a well-developed technology it provides
an invaluable test bed for small quantum computations.  
Optical schemes based on entangled photons from down-conversion
have also been used to implement a variety of small quantum
simulations, but since photons don't normally interact with each other,
they don't provide a natural route to special purpose quantum simulators.
We describe the lessons learned from these quantum simulations
in section \ref{proof}.
We then turn to simulators built by trapping arrays of ions, atoms,
and electrons in sections \ref{OL} and \ref{OA}.
Most of these have applications both as universal quantum simulators and
for specific Hamiltonians, with promising experiments and rapid 
progress being made with a number of specific configurations.

%% file: Proof.tex
\section{Proof-of-principle experiments}\label{proof}

Some of the most advanced experimental tests of quantum computation
have been performed using technology that does not scale up
beyond ten or so qubits.  Nonetheless, the information gained
from these experiments is invaluable for developing more scalable
architectures.  Many of the control techniques are directly transferable
in the form of carefully crafted pulse sequences with enhanced
resilience to errors and imperfections. 
Observing the actual effects of decoherence on the fidelities
is useful to increase our understanding of the requirements for scaling up
to longer sequences of operations.

\subsection{NMR experiments} \label{NMR}

Nuclear Magnetic Resonance is a highly developed technology
that provides an adaptable toy system for quantum computing
(see Jones \cite{Jones2000} for a comprehensive review).
A suitable molecule with atoms having various nuclear spins is prepared,
often requiring chemical synthesis to substitute different
isotopes with the required spins.  A solution of this
molecule then provides an ensemble which can be collectively controlled
by applied magnetic fields and radio frequency (rf) pulses.  The nuclear spins
of the different atomic species will in general have
different resonant frequencies, allowing them to be addressed
separately.  Read out is provided by exploiting spin echo effects.
Liquid state NMR isn't considered to be scalable
due to the difficulty of addressing individual qubits in larger molecules.
Nonetheless, the relative ease with which quantum algorithms can
be implemented for small systems has meant that many proof-of-principle
experiments have been carried out using NMR.
These are often of only the smallest non-trivial size, using as few as one or
two qubits, but are still useful for developing and testing the
control sequences.  The real advantage lies in the
flexibility of applying gates through radio frequency (RF) pulses.
This allows NMR to outperform other test-bed systems such as optics,
where each gate requires its own carefully aligned components on the bench.
Since most quantum algorithms have been tested in NMR by now, we
select for discussion a few that bring out important points about
the experimental feasibility of quantum simulation in general.

Numerous groups have performed NMR quantum simulations of spin
chains.  The Heisenberg interaction is already present in NMR in the form of 
the $ZZ$ interaction ($X$, $Y$, $Z$ are used to denote the Pauli
spin operators).  This allows more complex Heisenberg interactions
to be simulated by using local unitary operations to rotate the
spin between the $X$, $Y$ and $Z$ orientations.
These simulations are thus a simple example of using one fixed
Hamiltonian -- $ZZ$ in this case -- to simulate another, as
described theoretically in section \ref{Universal}.
This allows the 
investigation of interesting properties of these spin chains such as 
phase transitions \cite{Peng2005, Peng2009}, the propagation of excitons 
\cite{Khitrin2001} and the evolution under particular interactions 
\cite{Zhang2005}. 
Peng et al \cite{Peng2005} and Khitrin et al \cite{Khitrin2001} found
that the decoherence time of the system 
is often too short to get meaningful results, even for these small simulations.
The limited decoherence times were turned into an advantage by Alvarez et al 
\cite{Alvarez2010}, to study the effects on quantum information
transfer in spin chains.  As expected, they were able to show that
decoherence limits the distance over which quantum information
can be transferred, as well as limiting the time for which
it can be transferred.  This is an example of using the noise naturally
present in the quantum computer to simulate the effects on the
system under study, as described in section \ref{open}.

Tseng et al \cite{Tseng1999} describe how to simulate a general
three-body interaction using only the $ZZ$ interaction present in NMR,
and experimentally demonstrated a $ZZZ$ interaction.  This provided proof 
of principle for extending the repertoire of NMR quantum simulation
beyond two-body Hamiltonians, later comprehensively generalised
theoretically by D\"ur et al \cite{dur2007} (see section \ref{UniversalH}).
Liu et al \cite{Liu2008} demonstrated experimentally that four-body
interactions in a four qubit NMR quantum computer can be
simulated to within good agreement of their theoretical calculations.

Pairing Hamiltonians (see section \ref{BCS}) are of particular
importance for quantum simulation,
with the fermionic systems they describe including superconductors and 
atomic nuclei.  The long range interactions put simulation
of general pairing systems beyond the reach of classical computers.
Studies using NMR have focused on the BCS Hamiltonian,
equation (\ref{eq:BCS}),
which is a pairing Hamiltonian with interactions composed of Pauli spin
operators.
However, because it consists of two non-commuting parts, 
these have to be implemented individually and then recombined
using the Trotter-Suzuki formula, as described in section \ref{Lloyd}.
Wu et al. \cite{Wu2002} provided a detailed discussion of how to
make this efficient for NMR, and their method was implemented
experimentally by Brown et al \cite{Brown2006} on three qubits.
As well as their insightful comments on the scaling of errors 
in the simulation, discussed in section \ref{noise},
they also added artificial noise to their simulations
to verify the scaling.
This confirmed that simulation of larger systems will be
challenging, due to the high number of operations required for
the Trotter expansion, and correspondingly large error correction overheads.

Negrevergne et al \cite{Negrevergne2005} 
simulate a many-body Fermi system that obeys the Fano-Anderson model,
a ring system with an impurity at the centre. 
This can be done with three NMR qubits, once the translational symmetry
in the ring has been taken into account and the fermion modes mapped
to the qubits.
To minimise problems with decoherence caused by running the system for a long 
time, Negrevergne designed and implemented an approximate refocusing scheme.
This provided a scalable algorithm, which can be adapted to 
other architectures as more powerful quantum simulators are built.

Although bosons are easier to simulate classically than fermions
(because they don't suffer from the ``sign problem'') for quantum
simulations they are harder, due to the unlimited size of
the Hilbert space.  The Hilbert space has to be artificially
truncated, and this limits the accuracy.
Simulations of a bosonic system have been carried out by Somaroo et al 
\cite{Somaroo1999}, who chose the truncated harmonic 
oscillator.  The limitations due to the truncation are quite significant
in a small NMR simulation, and scaling up would be difficult,
as a larger system would require small couplings within the
NMR simulator that would severely limit the time scale of the experiment.
As with other simulations, the decoherence time 
limits the duration of the experiment, which in this case corresponds to 
the number of periods of the oscillator which can be simulated.

Du et al \cite{Du2010} have simulated molecular hydrogen in order 
to obtain its ground state energy. To do this they use the algorithm 
presented by Aspuru-Guzik et al \cite{Aspuru-Guzik2005}, described
in section \ref{qchem}.
This is an important class of quantum simulations, because it turns
out to be more efficient in the quantum case to simulate the
dynamics exactly, instead of following the approximations used to
do these calculations classically.  They thus offer the possibility
of significant improvements for quantum chemistry, given a large
enough quantum computer.  With NMR systems, the simulations are
limited to hydrogen, and while the decomposition of the molecular 
evolution operator scales efficiently, Du et al  \cite{Du2010}
are not sure whether the same is true of their adiabatic state
preparation method.  Nonetheless, this is an important proof
of principle for the method and application.

\subsection{Photonic systems}

Linear optics, with qubits encoded in the photonic degrees of freedom,
are an attractive option for quantum computing due to the
relatively straightforward experimental requirements compared to
architectures requiring low temperatures and vacuum chambers.
The main difficulty is obtaining a suitable nonlinear interaction,
without which only regimes that can be simulated efficiently
classically can be reached.
Current experiments generally use less scalable techniques for
generating the nonlinear operation, such as using entangled
pairs of photons from down-conversion in nonlinear crystals,
or measurements with probabilistic outcomes,
so the experiment has to be repeated until it succeeds.

Lanyon et al \cite{Lanyon2010} used the algorithm presented by
Aspuru-Guzik et al \cite{Aspuru-Guzik2005} to simulate molecules.
The qubits were encoded in the polarisation of
single photons, with linear optical elements and a nonlinearity
obtained through projective measurements used to provide the
necessary control.
Ma et al \cite{Ma2010} used the polarisation states of four photons
to simulate a spin system of four spin$-1/2$ particles with arbitrary 
Heisenberg-type interactions between them.  They used
measurements to induce the interaction between the spins, and were able
to measure ground state energy and quantum correlations for the four
spins.

While photonic systems do not have an intrinsic Hamiltonian that is adaptable
for special purpose quantum simulation, they are expected to come
into their own as universal quantum computers.
There are strong proposals for scalable architectures
based on photonic systems \cite{Kok2007,OBrien2009} that can also be
exploited for quantum simulation.

%% file: OL.tex
\section{Atom trap and ion trap architectures}\label{OL}

Among the architectures for quantum computing predicted to be
the most scalable, qubits based on atoms or ions in trap systems
are strongly favoured \cite{Schaetz2007,Leibrandt2009}.
Locating the atoms or ions in a trap allows each qubit to be distinguished,
and in many cases individually controlled.
Review of the many designs that are under development is beyond the
scope of this article; while any design for a quantum computer
is also suitable for quantum simulation, we focus here on 
arrays of atoms or ions where the intrinsic coupling between
them can be exploited for quantum simulation.

Trapped ions form a Coulomb crystal due to their mutual repulsion,
which separates them sufficiently to allow individual addressing
by lasers.  Coupling between them can be achieved via the
vibrational modes of the trap, or mediated by the controlling lasers.
Atoms in optical lattices formed by counter-propagating laser
beams are one of the most promising recent developments.  Once
the problem of loading a single atom into each trap was overcome
by exploiting the Mott transition \cite{Greiner2002}, the road was
clear for developing applications to quantum computing and quantum 
simulation.
For comprehensive reviews of experimental trap advances, see
Wineland \cite{Wineland2009} for ion trapping, and Bloch et al
\cite{Bloch2008} for cold atoms.

Jan\'e et al \cite{Jane2003} consider quantum simulation using both 
neutral atoms in an optical lattice and ions stored in an array of 
micro traps. This allows them to compare the experimental resources 
required for each scheme, as well as assessing the feasibility 
of using them as a universal quantum simulator.
Atoms in optical lattices have the advantage that there is a
high degree of parallelism in the manipulation of the qubits.
The difficulty of individually addressing each atom,
due to the trap spacing being of the same order as the wavelength
of the control lasers, can be circumvented in several ways.
If the atoms are spaced more widely, so only every fifth or
tenth trap is used, for example, then individual laser addressing
can be achieved.  Applied fields that intersect at the target
atom can also be used to shift the energy levels such that only
the target atom is affected by the control laser.
Jan\'e et al conclude that both architectures should
be suitable for quantum simulation.

An alternative approach is to avoid addressing individual atoms altogether.
Kraus et al \cite{Kraus2007} explore the potential of simulations using 
only global single-particle and nearest neighbor interactions. This 
is a good approximation for atoms in optical lattices, and
the three types of subsystem they consider -- fermions, bosons, and spins --
can be realised by choosing different atoms to trap in the optical lattice
and tuning the lattice parameters to different regimes.
They make the physically reasonable assumption that the interactions
are short range and translationally invariant.
They also apply an additional constraint of periodic boundary conditions,
to simplify the analysis.
Most physical systems have open rather than periodic boundary 
conditions, so their results may not be immediately applicable to
experiments.
For a quadratic Hamiltonian acting on fermions or bosons in a cubic lattice,
Kraus et al found that generic nearest neighbor interactions are 
universal for simulating any translationally invariant interaction when 
combined with all on-site Hamiltonians (the equivalent of any local unitary)
provided the interactions acted along both the axes and diagonals of the
cubic lattice (compare lattice gases, section \ref{LGA}).
However, for spins in a cubic lattice,
there is no set of nearest-neighbor interactions which is universal and 
not even all next-to-nearest neighbor interactions could be simulated 
from nearest-neighbor interactions.
It is possible that different encodings to those used by Kraus et al
could get around this restriction, but the full capabilities of
spin systems on a cubic lattice remains an open problem.
Their results demonstrate that schemes which don't provide 
individual addressability
can still be useful for simulating a large class of Hamiltonians.

Coupled cavity arrays are a more recent development, combining
the advantages a cavity confers in controlling an atom with
the scalability of micro-fabricated arrays.  While
there is a trade off between the relative advantages of the various 
available trapping architectures, with individual addressability and 
greater control resulting in systems with a poorer scaling in precision, 
each scheme has its own advantages and the experiments are still 
in the very early stages.

\subsection{Ion trap systems}\label{ion}

The greater degree of quantum control available for ions in traps,
compared with atoms in optical lattices,
means that research on using ion traps for simulating quantum
systems is further developed. 
Clark et al \cite{Clark2009} and Buluta and Hasegawa \cite{Buluta2009a} 
present designs based on planar RF traps that are specifically
geared towards quantum simulations.
They focus on producing a square lattice of trapped ions,
but their results can be generalised to other 
shapes such as hexagonal lattices (useful for studying systems such
as magnetic frustration).
Clark et al carried out experimental tests
on single traps that allowed them to verify their
numerical models of the scheme are accurate.  
They identify a possible difficulty 
when it is scaled to smaller ion-ion distances.
As the ion spacing decreases, the secular frequency increases,
which may make it difficult to achieve coupling strengths that are large
relative to the decoherence rate.

As with the simulations done with NMR computers,
some of the earliest work on ion
trap simulators has focused on the simulation of spin systems.
Deng et al \cite{Deng2005} and
Porras and Cirac \cite{Porras2004,Porras2004a} discuss the application
of trapped ions to simulate the Bose-Hubbard model, and Ising and
Heisenberg interactions.  This would allow the observation and
analysis of the
quantum phase transitions which occur in these systems.
They mention three different method for 
trapping ions that could be used to implement their simulation schemes. 
Arrays of micro ion traps and linear Paul traps use similar 
experimental configurations, although Paul traps allow a long range 
interaction that micro ion trap arrays don't.
Both schemes are particularly suited to simulating an interaction
of the form XYZ.
Penning traps containing two-dimensional Coulomb crystals
could also be used, and this would allow 
hexagonal lattices  to be applied to more complex simulations, such as 
magnetic frustration.
Alternatively \cite{Porras2004a}, the phonons in
the trapped ions can be viewed as the system for the simulation.
Within the ion trap system phonons can neither be created nor destroyed,
so it is possible to simulate systems such 
as Bose-Einstein condensates, which is more difficult using qubit systems. 

Friedenauer et al \cite{Friedenauer2008} have experimentally simulated a 
quantum phase transition in a spin system using two trapped ions.
The system adiabatically traverses from the quantum paramagnetic regime
to the quantum (anti)-ferromagnetic regime, with all the parameters
controlled using lasers and RF fields.  To extract data over the full
parameter range the experiment was repeated $10^4$ times, to obtain good
statistics for the probability distributions.  While the simulation
method is scalable, involving global application of the control fields,
it isn't clear the data extraction methods are practical for
larger simulations.  This work is significant for being one of the few
detailed proof-of-concept experimental studies done in a system
other than NMR, and demonstrates the progress made
in developing other architectures.
In Gerritsma et al \cite{gerritsma10a}, they simulate the Klein paradox,
in which electrons tunnel more easily through higher barriers than low
ones, by precisely tuning the parameters in their trapped ion system.
Edwards et al \cite{edwards10a} have simulated an Ising system with a
transverse field using three trapped ions.  They alter the Hamiltonian
adiabatically to study a wide range of ground state parameters,
thereby mapping out the magnetic phase diagram.  This system is
scalable up to many tens of ions, which would reach regimes currently
inaccessible to classical computation, allowing behavior towards
the thermodynamic limit to be studied in detail for general and
inhomogeneous spin systems.

Proof-of-principle simulations have also been done with single ions.
While less interesting than coupled ions, because the coupled systems
are where the Hilbert space scaling really favours quantum simulations,
these still test the controls and encoding required.
For example, Gerritsma et al \cite{Gerritsma2010} simulated the
Dirac equation using a single trapped ion, to model a relativistic 
quantum particle. The high level of control the ion trap provides allows 
information about regimes and effects that are difficult to simulate 
classically such as Zitterbewegung.

\subsection{Atoms in optical lattices}\label{atomOL}

Atoms trapped in the standing waves created by counter-propagating
lasers are one of the most exciting recent developments in quantum
computing architectures.  Their potential for the quantum simulation
of many-body systems was obvious from the beginning, and has been 
studied by many groups since the initial work of Jan\'e et al
\cite{Jane2003}.  Trotzky et al \cite{Trotzky2009} compare 
optical lattice experimental data with their own
classical Monte Carlo simulations, to validate the optical lattice
as a reliable model for quantum simulations of ultra-cold
strongly interacting Bose gases.  They find good agreement
for system sizes up to the limit of their simulations of $3\times 10^5$
particles.

The most promising way to use atoms in optical lattices for quantum simulation 
is as a special purpose simulator, taking advantage of the natural
interactions between the atoms.  This will allow larger
systems to be simulated well before this becomes possible with
universal quantum computers.
The following three examples illustrate the potential for thinking
creatively when looking for the best methods
to simulate difficult systems or regimes.  
Johnson et al \cite{Johnson2009} discuss the natural occurrence of
effective three-body and higher order interactions in two-body collisions
between atoms in optical lattices.  They use these to explain 
experimental results showing higher-than-expected decoherence rates.
Tuning these many-body interactions could be done using
Feshbach resonance control or manipulating the lattice potentials,
allowing them to be used for the simulation of effective field theories,
see section \ref{LGA}.
Ho et al \cite{Ho2009} propose that simulating the repulsive 
Hubbard model is best done using the attractive Hubbard model,
which should be easier to access experimentally.
Mapping between different regimes in the same model should
be simpler to implement, allowing access to 
results that are usually difficult experimentally.
As with the trapped ion schemes, one of the most common subjects
for simulation is many-body quantum phase transitions.
Kinoshita et al \cite{Kinoshita2004} use rubidium-87 atoms trapped
in a combination of two light traps.  By altering the trap strengths,
the interactions between the atoms can be controlled, allowing them to
behave like a one-dimensional Tonks-Girardeau gas through to a Bose-Einstein
condensate.
They find very good agreement with theoretical predictions for a 1D
Bose gas.
This is a good example of a special purpose simulator, since there are no
individual controls on the atoms, allowing only regimes dictated by the
globally controlled coupling to be realised.

\subsection{Atoms in coupled cavity arrays}\label{atomCCA}

Optical lattices are not the only way to trap arrays of atoms.
Coupled cavity arrays offer control over individual atoms
much more conveniently than with optical lattices.
In coupled cavities the qubits are represented by either
polaritons or hyperfine ground state levels, with the former allowing
continuous control, and the latter individual addressability.
The cavities themselves are an artificial system grown on a
microchip in which the qubits on the chip interact with the
field mode of the cavity, and the cavities are coupled by the
exchange of photons.
A simulation of the Heisenberg 
model is generally one of the earliest proof-of-principle simulations for
a new architecture, and Cho et al \cite{Cho2008} propose a technique to allow 
these coupled cavity arrays to do this.
Their method should apply generally to different
physical implementations of micro cavities.
Kay et al \cite{Kay2008} and Chen et al \cite{Chen2010} both 
discuss implementation of the Heisenberg model in specific
coupled cavity architectures.  They confirm that control 
over nearest and next-nearest neighbour coupling can be achieved,
but without short control pulses only global controls are available.
Schemes that give individual addressability
need short control pulses 
to modify the intrinsic interactions. These may necessitate
the use of the Trotter approximation,
making it more difficult to obtain high precision results in cavity arrays.
Ivanov et al \cite{Ivanov2009} look at exploiting the polaritons
in couple cavity arrays to simulate phase transitions, 
in the same way as Porras and Cirac \cite{Porras2004a} consider
using phonons in ion traps.
These proposals show the versatility and potential of coupled cavity arrays
for further development.

%% file: Electrons.tex
\section{Electrons and excitons} \label{OA}

While atoms and ions in arrays of traps are the most promising
scalable architectures for quantum simulation at present, electrons
can also be controlled and trapped suitably for quantum simulation.
This can be done either by confining free electrons, or
exploiting the electrons-hole pairs in quantum dots.
Superconducting qubits harness collective states of electrons
or quantized flux to form qubits from superconducting
circuits with Josephson junctions. We briefly describe 
applications of these architectures to quantum simulations that
exploit their special features.

\subsection{Spin lattices}\label{electrons}

Spin lattices are arrays of electrons, where the spin of the electron is
used as the qubit.  Persuading the electrons to line up in
the required configuration can be done in various ways.
Mostame and Sch\"utzhold \cite{Mostame2008}
propose to trap electrons using pairs of gold 
spheres attached to a silicon substrate
under a thin film of helium. The electrons float on the surface
of the helium and induce a charge on the spheres, which generates a double well 
potential and hence traps the electrons.
Mostame and Sch\"utzhold describe how to use this architecture to
simulate an Ising spin chain, from which the generalisation to
more complicated models can easily be made.
This model for trapping electrons is suggested to be
more scalable than atom or ion traps.  However, it may be 
difficult to realise experimentally,
due to the precise controls needed, particularly in 
the thickness of the film of helium.
Byrnes et al \cite{Byrnes2007} propose to confine a 2D electron gas 
using surface acoustic waves to create an `egg-carton' potential.
The advantage that this system has over optical lattices is that it 
produces long range interactions.  It should therefore be more suitable 
for simulating Hubbard dynamics, which originate from the long range 
Coulomb interaction.  This scheme will 
allow observations of quantum phase transitions in systems of strongly 
correlated electrons as well as the study of the
metal-insulator transition.

\subsection{Quantum dots}\label{qdots}

The trapped electrons or holes in a semiconductor quantum dot
can be exploited as qubits, with control provided via gate electrodes
or optical fields.
Instead of focusing on just the qubit degrees of freedom,
the whole quantum dot can be thought of as an
artificial atom, which may thus make them suitable to
simulate chemical reactions.
Quantum dots are now easy to make; the problem is to control their
parameters and location so they can be used collectively in a
predictable manner.
Smirnov et al \cite{Smirnov2007} discuss using the coupling of
quantum dots to model bond formation.
They consider one of the simplest possible systems for 
proof of principle calculations, the interaction $H + 
H_{2} \to H_{2} + H$, where the molecular bond between a pair
of hydrogen atoms switches to a different pair.
This can be simulated with a system of three 
coupled quantum dots, such as has been 
demonstrated experimentally \cite{Vidan2004, Gaudreau2006}.
The high level of control in quantum dot systems
will allow the detailed study of chemical reactions in 
conditions not available in real molecules.

\subsection{Superconducting architectures}\label{jj}

Superconducting architectures have been developing steadily
although in general they are a few years behind the atom and
ion trap systems.  As universal quantum computers they are
equally suitable in principle for quantum simulation.
Charge, phase and flux qubits can be constructed using Josephson
junction superconducting circuits, with controls provided by
a variety of externally applied fields.

An ingenious proposal from Pritchard et al \cite{pritchard10a}
describes how to use a systems of Josephson junctions for
simulation of molecular collisions.  The simulations
are restricted to the single excitation subspace of
an $n-$qubit system, which requires only an $n\times n$-dimensional
Hamiltonian.  In return for this subspace restriction,
the individual parameters in the Hamiltonian can be varied independently,
providing a high level of generality to the simulation.
They use a time-dependent rescaling of time to optimise the actual
run time of the simulation to minimise decoherence effects.
They test their method in an experiment with three tunable coupled
phase qubits simulating a three-channel molecular collision between
Na and He.  They study the fidelities achieved, and determine
the relationship between the fidelity and length of time the
simulation is run for.  Higher fidelities require longer
simulation times, but this is independent of $n$, showing this
aspect of the method is fully scalable.

%% file: Outlook.tex
\section{Outlook}

Quantum simulation is one of the primary short- to mid-term goals of
many research groups focusing on quantum computation.
The potential advances that even a modest quantum simulator
would unleash are substantial in a broad range of fields,
from material science (high temperature superconductors 
and magnetic materials) to quantum chemistry.
Quantum simulations are particular promising for
simulating fermionic many-body systems and their phase transitions,
where the ``sign problem'' limits efficient classical
numerical approximation techniques.
Larger quantum simulators could tackle problems in 
lattice QCD that currently consume a sizable fraction of
scientific computing power, while quantum simulations of
quantum chemistry have wide-ranging applications reaching as
far as the design of molecules for new drugs.
We have seen that the theoretical foundations have been 
laid quite comprehensively, providing detailed
methods for efficient quantum simulators, and calculations
that confirm their viability.

One significant issue that remains to be fully addressed is
the precision requirements for larger scale quantum
simulations.  Due to the one-to-one mapping between the
Hilbert space of the system and the Hilbert space of the
quantum simulator, the resources required for a given precision
scale inversely with the precision.  Compared with
digital (classical and qubit) computations, this is
exponentially more costly.  When combined with the
long control sequences required by Trotterization, this
threatens the viability of such simulations of even fairly modest
size.  

Special purpose quantum simulators designed with
similar Hamiltonians to the quantum system being studied
are the front runners for actually performing a useful
calculation beyond the reach of conventional computers.
These come in many forms, matching the variety of
common Hamiltonians describing physical systems.
Among the most developed and versatile, ion traps and atoms in
optical lattices are currently in the lead, although
micro-fabrication techniques are allowing more
sophisticated solid state trap arrays to catch up fast.
Actual experimental systems capable of
quantum simulations of a significant size are still in 
the future, but the designs and proof-of-concept experiments
already on the table provide a strong base from which to progress
on this exciting challenge.